\documentclass[preprints,article,accept,moreauthors,pdftex,10pt,a4paper]{mdpi} 

\usepackage{subfig}

\firstpage{1} 
\pubvolume{xx}
\issuenum{1}
\articlenumber{5}
\pubyear{2018}
\copyrightyear{2018}
\history{Received: date; Accepted: date; Published: date}

\Title{Chaotic dynamics in a quantum Fermi-Pasta-Ulam  problem}

\Author{Alexander L. Burin $^{1}$\orcidA{0000-0003-1922-1711}, Andrii O. Maksymov $^{1}$, Ma'ayan Schmidt $^{1}$ and  Il'ya Ya. Polishchuk  $^{2}$}

\AuthorNames{Alexander L. Burin, Andrii O. Maksymov, Ma'ayan Schmidt and Il'ya Ya. Polishchuk}

\address{%
$^{1}$ \quad Department of Chemistry, Tulane University, New Orleans LA 70118, USA; aburin@tulane.edu\\
$^{2}$ \quad National Reserach Center "Kurchatov Institute," 123182, Moscow, Russia; Moscow Institute of Physics and Technology, Dolgoprudnii, Moscw Reg, 141700, Russia; iyppolishchuk@gmail.com}

\corres{Correspondence: aburin@tulane.edu; Tel.: +1-504-296-5825}

\abstract{We investigate the emergence of chaotic dynamics in a quantum Fermi - Pasta - Ulam  problem  for anharmonic vibrations in atomic chains applying semi-quantitative analysis of resonant interactions complemented by exact diagonalization numerical studies.  The crossover energy separating chaotic high energy phase and localized (integrable) low energy phase is estimated. It decreases inversely proportionally to the number of atoms until approaching the quantum regime where this dependence saturates. The chaotic behavior appears at lower energies in systems with free or fixed ends boundary conditions compared to periodic systems. The applications of the theory to realistic molecules are discussed.}


\begin{document}

\setcounter{section}{-1} 
\section{Introduction}

Understanding vibrational energy flow in molecules is one of the challenges in modern science and technology \cite{AbeScience,NitzanScience03}. Vibrational energy flows control energetics 
of chemical reactions, determine heat balance in modern nano-devices \cite{AbeScience,SegalNitzan03,LeitnerBook,15LeitnerReview,LeitnerReviewProtein18} and can be manipulated similarly to electrons and photons and used to carry and process quantum information  \cite{RevModPhys12,Gruebele07,Gruebele14}. Intramolecular energy relaxation and transport are dramatically sensitive to the molecule's  ability to attain the thermal equilibrium \cite{LeitnerBook,Leitner18Entropy}.

After seminal work by Stewart and McDonalds \cite{Stewart83}, it has been realized  that   the internal vibrational relaxation can be absent or proceed very slowly in small enough  molecules and/or  at low temperature.  Based on these observations the concept of localization of low energy anharmonic vibrational states of poly-atomic molecules within the manifold of harmonic product states of almost independent normal modes was put forward by Logan and Wolynes  \cite{LoganWolynes90}. In earlier \cite{KaganMaksimov85,ab89TLSJETP,ab90Kontor} and later \cite{Shepelyansky97QD,Basko06,Gornyi05} work similar ideas have been developed for particle and spin systems. Theory was further extended combining  random matrix theory methods \cite{Leitner96,LW2,LW3}  and Bose Statistics
Triangle Rule approach \cite{BG1,BG2,BG3} and this extension was reasonably consistent with the experimental observations \cite{Stewart83}. 

This development is qualitatively consistent with the investigations of the classical counterpart problem of anharmonic vibrational dynamics. Its simplest realization in  atomic chains probed as a modeling system for irreversible dynamics was considered in the celebrated work by Fermi, Pasta and Ulam \cite{FPUclassic} (FPU), where the quasi-periodic behavior has been discovered for the evolution of the initial excitation instead of irreversible energy equipartition. In spite of over sixty years of investigations of the FPU problem, its complete understanding still remains a challenge \cite{BermanIsrailevFPUReview05,Lvov15,
Benettin18betaFPULyapunovExp,Flach17FPU}. 

Both quantum and classical  non-linear vibrational dynamics can be characterized by  a  critical energy separating low energy integrable (localized) and high energy chaotic (delocalized) behaviors.  In the chaotic regime each part of the system can be thermalized due to its interaction with the rest suggesting ergodic behavior in classical regime which is expressed by  the  eigenstate thermalization hypothesis \cite{SrednickiETH94,Deutsch91} in quantum regime. The position of the crossover energy separating two regimes determines the localization threshold. The threshold energy can be redefined in terms of the critical temperature corresponding to that energy. 

The knowledge of the localization threshold for an individual molecule is significant since the vibrational relaxation changes dramatically depending on whether the energy of the molecule is lower or higher than the threshold \cite{PandeyLeitner2016ThermSign,15LeitnerReview,Leitner18Entropy,ab10Igor}. In the latter case the vibrational relaxation follows standard Fermi Golden rule kinetics \cite{Cohen13}, while in the localized regime it is much slower. Therefore, the present work is focused on the localization threshold and its dependence on system size (number of atoms) and the strength of anharmonic interaction.

Since the properties of a molecule can be sensitive to its shape the consideration is restricted to the simple linear chain of atoms coupled by anharmonic interactions identical to the FPU problem \cite{FPUclassic}. This problem is relevant for the energy relaxation and transport in polymer chains used in the modern heat conducting devices  \cite{SegalNitzan03,SegalReview2016,PandeyLeitner17ThermPEGs,
PandeyLeitner2016ThermSign}. The anomalous increase of a thermal conductivity there with the system size  suggests a very slow thermalization or even the lack of one  \cite{PandeyLeitner17ThermPEGs}. The results for the FPU problem can be qualitatively relevant for the analysis of more complicated molecules.

The consideration is restricted to quantum mechanical systems. It has been suggested that the threshold energy separating localized and chaotic states decreases with the system size \cite{Chirikov59,IsraelovChirikov66,BermanKolovskii84FPU,Shepelyansky97,
Benettin18betaFPULyapunovExp,Lvov15,Flach17FPU}. This leads to the reduction of thermal energy below the vibrational quantization energy, which makes quantum effects inevitably significant for sufficiently large molecules. 

The paper is organized as follows. The FPU problems with different boundary conditions are formulated and briefly discussed in Sec. \ref{sec:Model}. The analysis of localization is performed combining analytical (Sec. \ref{sec:Reson})  and numerical (Sec. \ref{sec:num}) approaches  for the FPU problems with different boundary conditions. Both approaches  are reasonably consistent with each other and led to the predictions of analytical dependencies of localization threshold on system parameters that are discussed  in Sec. \ref{sec:disc} for organic molecules. The methods and brief conclusions are formulated in Secs. \ref{sec:Methods}, \ref{sec:concl}.

\section{Model} 
\label{sec:Model}

\begin{figure}[h!]
\centering
\subfloat[]{\includegraphics[scale=0.18]{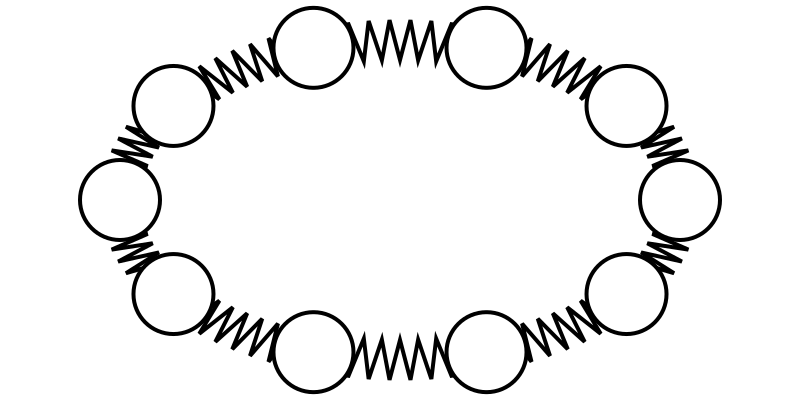}}
\subfloat[]{\includegraphics[scale=0.18]{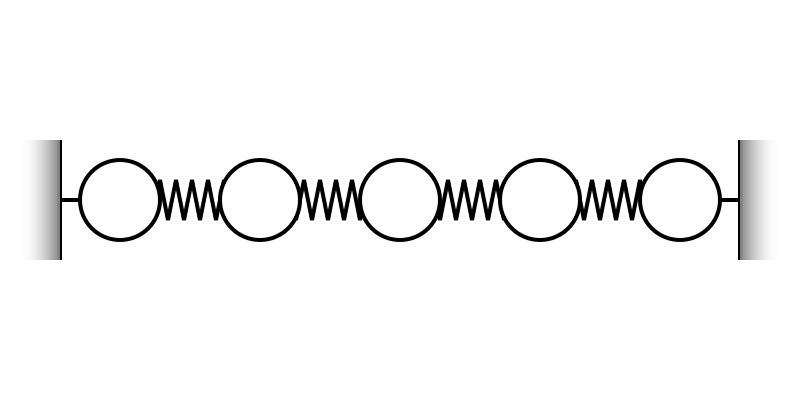}}
\subfloat[]{\includegraphics[scale=0.18]{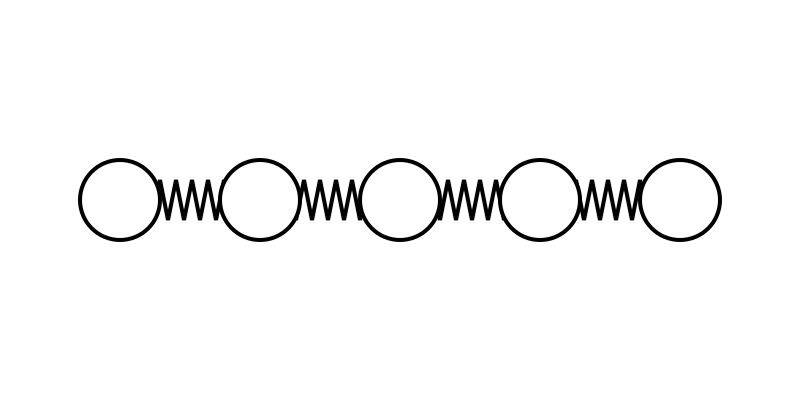}}
{\caption{Schematic illustration of periodic (a), fixed ends (b) and free ends (c) FPU atomic chains.}
\label{fig:Bound}}
\end{figure}

The FPU model of anharmonic atomic chain with different 
common boundary conditions including periodic, fixed ends and free ends (see Fig. \ref{fig:Bound}) can be described by the Hamiltonians defined as
\begin{eqnarray}
\widehat{H}_{per}=\sum_{i=1}^{N}\frac{\widehat{p}_{i}^2}{2M}+\sum_{i=1}^{N-1}\left[k\frac{(\widehat{u}_{i}-\widehat{u}_{i+1})^2}{2}+A\frac{(\widehat{u}_{i}-\widehat{u}_{i+1})^3}{6}+B\frac{(\widehat{u}_{i}-\widehat{u}_{i+1})^4}{24}\right] 
\nonumber\\
+k\frac{(\widehat{u}_{N}-\widehat{u}_{1})^2}{2}+A\frac{(\widehat{u}_{N}-\widehat{u}_{1})^3}{6}+B\frac{(\widehat{u}_{N}-\widehat{u}_{1})^4}{24}, ~ 
\text{periodic},
\nonumber\\
\widehat{H}_{fixed}=\sum_{i=1}^{N-1}\frac{\widehat{p}_{i}^2}{2M}+\sum_{i=1}^{N-2}\left[k\frac{(\widehat{u}_{i}-\widehat{u}_{i+1})^2}{2} + A\frac{(\widehat{u}_{i}-\widehat{u}_{i+1})^3}{6} +B\frac{(\widehat{u}_{i}-\widehat{u}_{i+1})^4}{24}\right]
\nonumber\\
+k\frac{\widehat{u}_{1}^2+\widehat{u}_{N-1}^2}{2}+A\frac{\widehat{u}_{1}^3+\widehat{u}_{N-1}^3}{6}+B\frac{\widehat{u}_{1}^4+\widehat{u}_{N-1}^4}{24}, ~ 
\text{fixed ends},
\nonumber\\
\widehat{H}_{free}=\sum_{i=1}^{N}\frac{\widehat{p}_{i}^2}{2M}+\sum_{i=1}^{N-1}\left[k\frac{(\widehat{u}_{i}-\widehat{u}_{i+1})^2}{2}+A\frac{(\widehat{u}_{i}-\widehat{u}_{i+1})^3}{6} +B\frac{(\widehat{u}_{i}-\widehat{u}_{i+1})^4}{24}\right], ~ \text{free ends}. 
\label{eq:H}
\end{eqnarray} 
Below we set mass, harmonic force constant and Planck constant to unity $\hbar=M=k=1$. Force constants $A$ and $B$ describe relative strengths of third and fourth order anharmonic interactions. The fixed ends problem has been studied in the classical FPU paper \cite{FPUclassic}.

Anharmonic interactions should be weak at the system energy $E$ of interest  to justify the applicability of the series expansion for non-linear terms. Assuming approximate energy equipartition  one can estimate $(x_{i}-x_{i+1})^2 \sim E/N$, which leads to anharmonic interaction estimates $V_{3} \sim AE^{3/2}/\sqrt{N}$ and $V_{4} \sim BE^{2}/N$ for the third and fourth order anharmonic interactions, respectively. Comparing harmonic and anharmonic interactions we end with the restrictions for energy density in the form 
\begin{eqnarray}
\frac{E}{N} < \frac{1}{B}, ~ \frac{1}{A^2}. 
\label{eq:EnDensIneq}
\end{eqnarray}


However we impose a stronger constraint on the energy requiring the stability with respect to the dissociation. Consequently, the total energy should be less than the dissociation energy, $E_{d}$. The dissociation energy can be estimated for the single bond assuming that anharmonic energy becomes comparable to the harmonic one which suggests $E_{d} \sim \frac{1}{A^2} \sim \frac{1}{B}$. For instance if for the Morse potential \cite{MorseLinnett1941} often used to model atomic interactions one has 
\begin{eqnarray}
E_{d} = \frac{7}{2B}= \frac{9}{4A^2}. 
\label{eq:DissEn}
\end{eqnarray} 
We assume that the system energy is always smaller than the dissociation energy 
\begin{eqnarray}
E < E_{d} \sim \frac{1}{B} \sim \frac{1}{A^2}, 
\label{eq:DissConstr}
\end{eqnarray} 
so the molecule is stable with respect to large coordinate displacements. 

Eq. (\ref{eq:DissConstr}) can be satisfied for a quantum system only if it is satisfied at least for the minimum energy that can be estimated as a quantization energy $E \sim 1$. Consequently, the anharmonic interactions should be weak that requires 
\begin{eqnarray}
A, B \ll 1.
\label{eq:weakAnh}
\end{eqnarray} 
These requirements are well satisfied in real molecules because of the small amplitudes of vibrations of heavy atoms. For instance using the Morse potential for C$-$C bond one can express the dimensionlass parameters $A$ and $B$ as 
\begin{eqnarray}
|A| = \frac{3\sqrt{2}}{2}\sqrt{\frac{\hbar \sqrt{k/M}}{E_{d}}}= 0.4644, ~ B =\frac{7}{2}\frac{\hbar \sqrt{k/M}}{E_{d}}=0.2644.
\label{eq:weakAnhMorseEst}
\end{eqnarray} 
The conditions of the weakness of anharmonicity are better satisfied  since the dissociation energy contains additional large numerical factor, see Eq. (\ref{eq:DissEn}). 

In the absence of anharmonic interactions one can describe periodic chain in terms of its normal modes that can be characterized by quasi momenta quantum numbers $p=-N/2+1, -N/2+2, ... N/2$ for even $N$ and $p=-(N-1)/2, -(N-3)/2,... (N-1)/2$ for odd $N$, wavefunctions $\psi_{p}(k)=e^{i2\pi pk/N}/\sqrt{N}$ and eigenfrequencies 
\begin{eqnarray}
\omega_{p}=2\sin(p\pi/N). 
\label{eq:Freq}
\end{eqnarray}
These frequencies are identical to normal mode quantization energies (remember that we set the Planck constant $\hbar$ to unity). 
The mode with zero quasi-momentum $p=0$ can be excluded from the consideration because it corresponds to identical displacements of all atoms that cannot modify a system energy. Therefore, only $N-1$ normal modes are significant and the harmonic part of the Hamiltonian can be reexpressed in the diagonal form with respect to   these normal modes as 
\begin{eqnarray}
\widehat{H}_{0} = \sum_{p=1}^{N-1}\omega_{p}\left[\widehat{b}_{p}^{\dagger}\widehat{b}_{p}+\frac{1}{2}\right], 
\label{eq:harmdiag}
\end{eqnarray}
where operators $\widehat{b}_{p}^{\dagger}$, $\widehat{b}_{p}$ describe creation or annihilation of one quantum of vibration of normal mode or phonon $p$. 
The harmonic problem is obviously integrable since the system breaks into $N$ independent oscillators (phonons) and each phonon population number operator $\nu_{p}=\widehat{b}_{p}^{\dagger}\widehat{b}_{p}$ represents a local integral of motion \cite{Huse14IntMot} in the momentum representation. Each many-body eigenstate $|S>$ of the harmonic system can be then represented by an arbitrarily sequence of integer population numbers $S=\{\nu_{p}\}$. 

Anharmonic interaction mixes up these states because it breaks down the conservation of individual phonon population numbers. In the periodic system with only fourth order anharmonicity ($A=0$ in Eq. (\ref{eq:H})) this interaction can be expressed as  
\begin{eqnarray}
\widehat{V}=\frac{B}{96N}\sum_{p_{1}p_{2}p_{3}p_{4}}\Delta(p_{1}+p_{2}+p_{3}+p_{4}){\rm sign}(p_{1}p_{2}p_{3}p_{4})(-1)^{\frac{(p_{1}+p_{2}+p_{3}+p_{4})}{N}}\sqrt{\omega_{p1}\omega_{p2}\omega_{p3}\omega_{p4}}\times 
\nonumber\\
\times (\widehat{b}^{\dagger}_{p1}+\widehat{b}_{-p1})(\widehat{b}^{\dagger}_{p2}+\widehat{b}_{-p2})(\widehat{b}^{\dagger}_{p3}+\widehat{b}_{-p3})(\widehat{b}^{\dagger}_{p4}+\widehat{b}_{-p4}), ~
\Delta(p)=\frac{1}{N}\sum_{k=1}^{N}e^{\frac{i2\pi pk}{N}}. 
\label{eq:anharm}
\end{eqnarray}
The factor $\Delta(p_{1}+p_{2}+p_{3}+p_{4})$ is equal to unity if the sum of all four momenta is equal to zero or integer fraction of $N$  (due to Unklamp processes); otherwise it is equal to zero giving rise to a quasi-momentum conservation. 

Because of the above conservation law the basis states of the system with given normal mode population numbers can be split into $N$ subsystems  with the total  quasi-momenta $Q=0, 1, -1, ...N/2$ for even $N$ or $Q=0, 1, -1, ...-(N-1)/2$ for odd $N$ determined with the accuracy to the addition of integer number of $N$'s. Each subsystem should be studied separately since the states from different subsystems do not interact with each other. In addition the states with $Q=0$ and $Q=N/2$ for even $N$ possess a mirror reflection symmetry with respect to replacement all  states $S={\nu_{p}}$ with the states $S_{-}={\nu_{-p}}$. Then the states with $Q=0, N/2$ can be split into two subgroups symmetric or antisymmetric with respect to the mirror reflection symmetry. Consequently, all many-body states can be split into $N+1$ subgroups for odd $N$ and $N+2$ subgroups for even $N$ that should be considered separately. 

Similarly, one can consider the interacting normal modes for free and fixed boundary atoms (see Eq. (\ref{eq:H})). One can similarly introduce normal modes for this problem and their anharmonic coupling.  In these two cases one cannot introduce quasi-momenta because of the lack of translational symmetry. Yet there is a mirror reflection symmetry with respect to the middle of the chain. Then all states can be separated into two subgroups of symmetric and anti-symmetric states with respect to that symmetry. The states belonging to different subgroups can be considered separately.

\section{Localization-delocalization transition: Qualitative analytical consideration}
\label{sec:Reson}

There is over sixty years history of the investigation of chaos in the classical FPU problem and this problem still remains a challenge \cite{BermanIsrailevFPUReview05,Lvov15}. The situation with the quantum mechanical problem is even more complicated \cite{SchickFPULuttLiq68,abFPUPreprint}. Below we summarize the established results for the system of a few atoms \cite{Cohen13} and attempt to extend them to atomic chains having many  atoms $N \gg 1$ first for the $\beta$ FPU problem containing only fourth order anharmonic interactions and then extend it to the mixed $\alpha +\beta$ problem. 

\subsection{Localization - Chaos transition in the small system $N \sim 1$.}
\label{sec:Small}

Localization-chaos transition in systems with the small number of atoms $N \sim 1å$ can be well understood following the Ref. \cite{Cohen13}. In this paper the critical energy $E_{c}$ separating localized and chaotic states  was estimated using a dimensionality arguments since the only value having the dimension of energy can be constructed using the parameters in  Eq. (\ref{eq:H}) in the form 
\begin{eqnarray}
E_{c} \sim \frac{1}{B}. 
\label{eq:EcSmallN}
\end{eqnarray} 
The numerical studies of both classical and quantum mechanical problems in Ref. \cite{Cohen13} have confirmed these expectations provided that the system is semiclassical, i. e. the system energy, $E_{c}$ is much larger than the quantization energy $\hbar\omega \sim 1$. This requires $B \ll 1$.  Transitions in quantum and classical systems occur under almost identical conditions  since the system is semiclassical because the maximum quantization energy max$(\omega_{p})$ is of order of unity (Eq. (\ref{eq:Freq}). Consequently, it is much less than the energy per the mode $E_{c}/N$ expressing the thermal energy $k_{B}T$ (see Eq. (\ref{eq:EcSmallN}), remember that $B \leq 1$ and $N \sim 1$). 

Similarly one can estimate the critical energy for the $\alpha$ FPU problem with the third order anharmonic interaction as 
\begin{eqnarray}
E_{c} \sim \frac{1}{A^2}. 
\label{eq:EcSmallNalph}
\end{eqnarray} 

Eqs. (\ref{eq:EcSmallN}), (\ref{eq:EcSmallNalph}) differ from the expectations of the analysis exploiting resonances for many-body transitions that has been successfully applied to  problems of interacting spins \cite{ab90Kontor,ab06preprint,ab15MBL,ab16preprintSG} or electrons \cite{Shepelyansky97QD,abGorniyMirlinDot}. According to this criterion chaos emerges in the presence of approximately one resonance per the many-body state under the condition that the diagonal interaction of resonant modes is larger or comparable to their resonant coupling   \cite{ab90Kontor,ab95TLSRelax,ab16preprintSG,abGorniyMirlinDot} that is needed to avoid destructive interference between consecutive resonant transitions. Such interaction is present naturally  for the fourth order anharmonicity in Eq. (\ref{eq:anharm})  (for instance, the terms with $p_{1}=p_{2}$ and $p_{3}=p_{4}$ are diagonal in the phonon product state representation). There is no such interaction in the case of the third order anharmonic interactions ($\alpha$ FPU problem), which changes the definition of delocalization transition as it is discussed in Sec. \ref{sec:AlphaRes}. 

However, in the $\beta$ FPU problem under consideration the matrix elements $M$ of the four phonon  interactions in Eq. (\ref{eq:anharm}) grow proportionally to the squared population numbers $M\sim B\nu_{p}^{2} \sim B E^2$ for the system energy $E$ exceeding the quantization energy. The typical energy change in a four phonon process is of order of  their quantization energy that is of order of unity. Consequently, the amount of resonances approaches unity at $E \sim B^{-1/2}$ in contrast with Eq. (\ref{eq:EcSmallN}). 

This conflict can be resolved at the qualitative level modifying the definition of resonances in accordance with Ref. \cite{Levitov90} where a single-particle localization problem has been considered for harmonically coupled vibrations. For example two unit mass oscillators with frequencies $\omega_{a}$ and $\omega_{b}$ coupled by the interaction $k_{ab}u_{a}u_{b}$ are in resonance under the condition $|\omega_{a}^2-\omega_{b}^2|<k_{ab}$, while in terms of matrix elements the resonance takes place at $|\omega_{a}^2-\omega_{b}^2|<k_{ab}\sqrt{(\nu_{a}+1)(\nu_{b}+1)}$. 

To define the resonance correctly one can consider the energy change not for a single resonant transition but for the whole set of possible transitions involving these four phonons, which will increase  the typical energy change due to the transition ($\hbar\omega \sim 1$) by the factor of a typical  phonon population number $\nu_{p}$. Then the resonance criterion can be written as $B\nu_{p}^{2} \sim \nu_{p}$. Setting $\delta E \sim \nu_{p} \approx E_{c}$ we end up with Eq. (\ref{eq:EcSmallN}). The problem of interest with large number of atoms needs a special consideration given in the next section.

\subsection{$\beta$ FPU problem}
\label{sec:BetaRes}


\subsubsection{Classical Regime}
\label{sec:betclass}

Consider the localization - chaos transition  in the case of a large number of atoms assuming that the system is semiclassical (population of each vibrational mode exceeds unity). One can still consider  resonant interactions  similarly to the previous section. In the periodic system of $N$ atoms one can find $N^{3}$ possible four phonon processes for a typical state (the fourth phonon mode is fixed by the quasi-momentum conservation law in Eq. (\ref{eq:anharm})). Consequently, the minimum energy difference between two modes coupled by the fourth order anharmonic interaction  is given by $\delta E \sim \nu_{p}/N^{3} \sim E/N^4$ (factor $\nu_{p}$ is added similarly to Sec. \ref{sec:Small}). The interaction matrix element scales as  $M \sim  B E^2/N^3$. Here the factor $1/N$ comes from the definition of anharmonic interactions in Eq. (\ref{eq:anharm}) and the factor $\nu_{p}^2 \sim (E/N)^2$ is determined by the population numbers $\sqrt{\nu_{p_{1}}\nu_{p_{1}}\nu_{p_{2}}\nu_{p_{4}}} \sim (k_{B}T)^2$ while the thermal energy $k_{B}T$ for $N$ classical oscillators is given by $E/N$ \cite{kittel2004introduction}. Setting $\delta E \sim M$ to ensure the presence of resonant interactions we estimate the localization threshold as 
\begin{eqnarray}
E_{c,res}(N) \sim \frac{1}{NB}. 
\label{eq:crossResEst}
\end{eqnarray}

Similar dependence can be obtained for the atomic chains with fixed or free ends boundary conditions where there is no quasi-momentum conservation. In those systems one has $N^{4}$ possible four phonon transitions and $1/N^2$ (instead of $1/N$, Eq. (\ref{eq:anharm})) scaling of anharmonic interaction matrix element. Then extra factors $N^{-1}$ are canceled out on both sides of the criterion of resonance leading to Eq. (\ref{eq:crossResEst}).

It is noticable that the estimated behavior of localization threshold in Eq. (\ref{eq:crossResEst})  is qualitatively consistent with the earlier estimates \cite{BermanKolovskii84FPU,Dauxois97FPU,Budinsky83FPU,Shepelyansky97} obtained using the stability analysis of the classical dynamics of a non-linear FPU chain in the form 
\begin{eqnarray}
E_{c}(N) \approx \frac{2\pi^2}{NB}. 
\label{eq:crossResEarlier}
\end{eqnarray}
Since this equation agrees with numerical studies in Sec. \ref{sec:num} for free and fixed ends boundary conditions we will use it for quantitative estimates.

In our qualitative analysis of resonant interactions we considered only typical phonons with energy close to unity, while the low frequency phonons were ignored. Based on the present understanding of localization - chaos transition it is hard to expect that they can suppress the chaotic dynamics because the typical phonons form the ergodic spot normally capable to equilibrate the rest of the system \cite{abGorniyMirlinDot,Huveneers16HotSpots}. It is hard to expect that they can give additional support to the chaotic dynamics since they are coupled weakly to the rest of the system compared to typical phonons. 

On the other hand one can imagine marginal states with the only low frequency phonons being excited. These states can possibly show anomalously strong localization behavior as predicted for the classical systems in of Refs,  \cite{IsraelovChirikov66,Shepelyansky98}.  There are other suggestions for classical systems   \cite{Pettini90,Benettin18betaFPULyapunovExp,Lvov15,Flach17FPU} 
that the crossover in Eq. (\ref{eq:crossResEarlier}) does not describe the  transition  to a truly integrable (localized) behavior but separates strongly ergodic and weakly ergodic regimes at high and low energies, respectively. Since the numerical simulations in Sec. \ref{sec:num} show the pure localization transition, we did not see any evidences for such behavior in a quantum regime. It cannot be excluded that at larger number of atoms some additional channels for chaotic behavior can emerge. 

The criterion in Eq. (\ref{eq:crossResEst}) is valid until the system remains semiclassical, meaning that the phonon population numbers exceed unity. This requires the thermal energy $E_{c}(N)/N$ to exceed the quantization energy, which is of order of unity. Thus, the classical regime takes place at sufficiently small number of atoms 
\begin{eqnarray}
N<N_{c}  \approx \frac{\sqrt{2} \pi}{\sqrt{B}}.
\label{eq:maxN}
\end{eqnarray}

The crossover energy $E_{c}$ expresses the minimum threshold energy in the classical regime of vibrations. At larger $N$ the system should be treated quantum mechanically as considered in Sec. \ref{sec:betquant}. 

\subsubsection{Quantum Mechanical Regime}
\label{sec:betquant}

We begin the consideration with the analysis of the problem in terms of resonant interactions. 
Imagine that the system energy is spread between phonons of energy $\epsilon$ such that  $\sqrt{E/N} < \epsilon \leq 1$. In our case of small energy $E<N$ the thermal energy is given by $k_{B}T \approx \sqrt{E/N}$ and the lower limit for the energy $\epsilon$ qualitatively represents the typical thermodynamic equilibrium. 

For an arbitrary energy $\epsilon$ the total number of phonons is given by $n_{\epsilon} \sim E/\epsilon$. This number  is smaller than the number of quantum states with energy of order of $\epsilon$ that is given by $N\epsilon$. The modeling system is non-degenerate so typical populations of vibrational states do not exceed unity and one can describe the emergence of chaos requiring a single resonant interaction per a many-body quantum state (cf. Refs.  \cite{abGorniyMirlinDot,ab16preprintSG}). The typical anharmonic interaction strength for periodic boundary conditions scales as $M \sim  B \epsilon^2/N$. The energy difference to the adjacent state coupled to the given state and having the same number of phonons can be estimated as $\epsilon/N_{c}$ where $N_{c}$ is the number of anharmonically coupled states with the same number of phonons. This number can be estimated considering the number of possible anharmonic transitions including  $n^2$ possible double annihilations of phonons and $N\epsilon$ creations (the fourth phonon is fixed by the quasi-momentum conservation law and we consider only processes conserving the number of phonons). The resonant interactions exist under the condition $B \epsilon^2/N < \epsilon/(n^2\epsilon N)$.  Then the critical energy $E_{c}=n\epsilon$ can be estimated as  
\begin{eqnarray}
E_{c} \sim \frac{1}{\sqrt{B}}. 
\label{eq:ResBelowDe}
\end{eqnarray}
The generalization to the non-periodic boundary conditions can be done similarly to that in the previous section. 

This answer is universal and insensitive to the number of atoms. It predicts the saturation of the dependence of critical energy on the number of atoms in the quantum regime. Based on the numerical results in Sec. \ref{sec:num} we assume that the saturation takes place at $E_{c} =N$. Then combining Eqs. (\ref{eq:crossResEarlier}) with Eq. (\ref{eq:ResBelowDe}) one can write the summary of the predicted behaviors as 
\begin{eqnarray}
E_{c}=
\begin{cases}
  \frac{2\pi^2}{NB}, & N < \frac{\sqrt{2} \pi}{\sqrt{B}}, \\    
  \frac{\sqrt{2} \pi}{\sqrt{B}}, &    N > \frac{\sqrt{2} \pi}{\sqrt{B}}.
\end{cases}
\label{eq:ANS}
\end{eqnarray}

The qualitative behavior predicted by Eq. (\ref{eq:ANS}) is obtained using resonant language similarly to Ref. \cite{ab16preprintSG}, where the matching Bethe lattice problem has been used to justify the results. Similarly to Bethe lattice problem one can expect the appearance of the additional logarithmic factor in Eq. (\ref{eq:ANS}). However in our specific case it is of order of $1$ since the argument of logarith is determined by the ratio of diagonal and off-diagonal interactions \cite{ab16preprintSG}, which have same order of magnitude in the problem under consideration. The quantitative expression in Eq. (\ref{eq:ANS}) gives a reasonable estimate by order of magnitude but does not pretend to be the accurate expression.

In Sec. \ref{sec:num} it is verified for the minimum division of energy $E$ into phonons with energies of order of $1$. The more accurate numerical analysis of the problem is postponed for future. 

The consideration ignores correlations between phonon energies and momenta, that can take place due to quasi-momentum conservation in a periodic system \cite{abFPUPreprint}   or some trace of its conservation in the system with fixed end boundary conditions. These processes are fully suppressed for free ends boundary conditions where the above consideration is most applicable. It is less applicable for the periodic system where these correlations can be significant. For very small system energies comparable to the maximum quantization energy $1$ the periodic system becomes integrable \cite{abFPUPreprint} so the consideration fails. We still believe that our consideration is valid even for a periodic system where the hot spot \cite{Huveneers16HotSpots} can be formed by several excited phonons with nearly maximum energy. These phonons, indeed, form chaotic state (see Sec. \ref{sec:num}) and can equilibrate other parts of the system. The accurate numerical verification should resolve the raised questions.

\subsection{$\alpha$ FPU problem}
\label{sec:AlphaRes}

Here we consider  the effect of the third order anharmonic interaction on the state of the system. Let us begin the consideration with  the classical regime, $E>N$. For the small number of atoms the dimension based arguments lead to the estimate $A_{c} \sim 1/\sqrt{E}$, Eq. (\ref{eq:EcSmallNalph}). For a large number of atoms $N$ in the periodic system one can find $N^{2}$ possible three phonon processes  so the minimum energy shift can be estimated as $\delta E \sim \nu_{p}/N^{2} \sim E/N^3$. Remember that the third phonon state is fixed by the quasi-momentum conservation law in Eq. (\ref{eq:anharm}). 
The interaction matrix element scales as  $M \sim  A E^{3/2}/N^2$. Consequently, there are resonant interactions in the case of sufficiently large energy $E>E_{3res}(N)$, where 
\begin{eqnarray}
E_{3res}(N) \sim \frac{1}{N^2A^2}. 
\label{eq:crossResEstalph}
\end{eqnarray}
This estimate is consistent with Ref. \cite{Budinsky83FPU}; yet we do not think it  describes the localization breakdown correctly because of the lack of the diagonal interaction. In this case resonant transitions are independent of each other \cite{abGorniyMirlinDot,ab16preprintSG} which prevents the system from delocalization similarly to the $XY$ model, where there is no diagonal interaction \cite{ab15MBLXY}. Following Ref. \cite{ab15MBLXY} one can consider the induced resonant interaction in higher orders anharmonicity following the Schrieffer and Wolff method \cite{SchriefferWolff66}. In the first non-vanishing  order  the fourth order anharmonic interaction will be generated. This generated interaction is  similar to the one in the $\beta$ FPU problem with the effective interaction constant $B_{*} \sim A^2$ if expressed in the momentum space.  However, the induced diagonal interaction is much less than the third order resonant interaction because $A \ll 1$ so it cannot enhance delocalization due to three phonon transitions. Following Refs. \cite{Gornyi05,Basko06} one can suggest the weaker delocalization criterion of one resonance per each normal mode. This leads to the criterion $E_{c3} \sim 1/A^2$ that is insensitive to the number of atoms. 

However, the more efficient delocalization should take place due to the induced fourth order interaction characterized by the interaction strength  $B_{*} \sim A^2$. In that case one can expect the chaotic behavior following the estimate of Eq. (\ref{eq:EcSmallN}) that reads 
\begin{eqnarray}
E_{c\alpha}(N) \sim \frac{1}{NA^2}. 
\label{eq:Ecalpha}
\end{eqnarray}
Similarly to Sec. \ref{sec:betquant} this criterion is valid in the classical regime realized at $N < 1/A$ while in the opposite regime this dependence saturates at 
\begin{eqnarray}
E_{c\alpha} \sim \frac{1}{\sqrt{A}}. 
\label{eq:Ecalph1}
\end{eqnarray}

Following Ref. \cite{Benettin18betaFPULyapunovExp} one can expect that this prediction should be valid to the same extent as  Eq. (\ref{eq:EcSmallN}). Indeed, if one considers the combined $\alpha+\beta$ problem containing both third and fourth order anharmonic interactions then the chaotic state formation is dramatically suppressed at $B=4A^2/9$ because under these conditions the non-linear interaction would be identical to power series expansion of the integrable Toda model \cite{Toda67}. Consequently, in this regime one can expect that the third order anharmonic interaction characterized by the constant $A$ should produce similar delocalization effect to  the fourth order problem, characterized by the interaction constant $B \sim A^2$ in a full accord with the estimate of Eq. (\ref{eq:Ecalpha}). 

Following the recipes of Ref. \cite{Benettin18betaFPULyapunovExp} one can extend the above  consideration to the general  $\alpha +\beta$ problem, which can be reduced to the $\beta$ FPU problem with the interaction constant $B_{*}$ defined as 
\begin{eqnarray}
B_{*}=B-\frac{4A^2}{9}. 
\label{eq:B*}
\end{eqnarray}
Consequently, one can predict the localization threshold energy for the $\alpha+\beta$ problem in the form of generalized Eq. (\ref{eq:ANS})  
\begin{eqnarray}
E_{c}=
\begin{cases}
  \frac{2\pi^2}{N\left(B-\frac{4A^2}{9}\right)}, & N < \frac{\sqrt{2} \pi}{\sqrt{B-\frac{4A^2}{9}}}, \\    
  \frac{\sqrt{2} \pi}{\sqrt{B-\frac{4A^2}{9}}}, &    N > \frac{\sqrt{2} \pi}{\sqrt{B-\frac{4A^2}{9}}}.
\end{cases}
\label{eq:ANSGen}
\end{eqnarray}
This result is not applicable if the denominator in Eq. (\ref{eq:ANSGen}) is very close to zero. In the  case of nearly zero denominator the problem can be effectively described by the sixth order anharmonic interaction 
with the interaction constant $C=B^2$ \cite{Benettin18betaFPULyapunovExp}. In this regime the similar analysis of resonant interactions can be applied leading to the threshold energy behaviors $E_{c} \sim 1/(B\sqrt{N})$ in the classical regime and $E_{c} \sim 1/B^{2/3}$ in the quantum regime, where $N>B^{-2/3}$. 


Eq. (\ref{eq:ANSGen}) is the main result of the present work. In the next section some numerical justification is given based on the diagonalization of Hamiltonians in Eq. (\ref{eq:H}) within the reduced basis of many-body states.

\section{Numerical Analysis of the Transition Localization - Chaos} 
\label{sec:num}

The numerical analysis is limited to the $\beta$ FPU problem to avoid overcomplexity. Below the numerical studies attempting to justify the analytical predictions of Sec. \ref{sec:Reson} are reported. In Sec. \ref{sec:Stat} we define the numerical criterion of the chaotic behavior. Since the basis of many-body states is infinitely large one should restrict the phase space. In Sec. \ref{sec:approx} we introduce the method of basis restriction considering the states with the fixed number of phonons. In Sec. \ref{sec:scaling} we investigate the dependence of the localization threshold on the system energy (number of phonons) and number of atoms. 

\subsection{Level Statistics} 
\label{sec:Stat}

The chaotic and integrable (or localized) phase of quantum systems can be identified using the statistics of energy levels. It is expected that in the chaotic phase all states substantially overlap with each other which leads to their energy level repulsion and, consequently, Wigner-Dyson level statistics \cite{ShklovskiiShapiro93,OganesyanHuse07} suggesting zero probability density for nearest eigenstates energy difference approaching zero. In the localized phase the overlaps of a majority of states are negligible so their energies are independent, which results in the Poisson statistics for energy level differences. In numerical studies exploiting exact diagonalization of the system Hamiltonian  the energy level statistics can be probed directly and used to identify the state of the system. 

Other methods including the analysis of correlation functions \cite{Yao14MBLLongRange,ab15MBL}, entanglement entropy \cite{Pollman12EntEntr} or local integrals of motion \cite{Huse14IntMot} can  also be used to study the delocalization with respect to the specific basis. However, the results depend on the choice of the basis. For instance, the basis of single particle states can be defined in the coordinate or momentum representations and localization in the coordinate space suggests delocalization in the momentum space and vice versa. Eigenstates of the FPU problem at very low energies \cite{abFPUPreprint} are delocalized in the basis of product states composed by independent phonon states, while the problem remains integrable \cite{SchickFPULuttLiq68}. The level statistics based definition is basis independent and therefore it seems to be the most objective criterion to distinguish localized and chaotic  phases. 

The level statistics have been characterized using the averaged ratio of successive gaps, $<r>$, defined as \cite{OganesyanHuse07} 
\begin{eqnarray}
<r> = \left<  \frac{{\rm min} (\delta_{n}, \delta_{n+1})}{{\rm max}(\delta_{n}, \delta_{n+1})} \right>,
\label{eq:Og}
\end{eqnarray}
where $\delta_{n} =E_{n+1}-E_{n}$ is the energy difference of adjacent energy levels of the system, Eq. (\ref{eq:H}), obtained by means of exact diagonalization of the Hamiltonian. According to Ref. \cite{OganesyanHuse07} in the chaotic regime characterized by Wigner-Dyson statistics  one has $<r>\approx 0.5307$, while in the case of localization where the Poisson statistics is expected one has $<r> \approx 0.3863$ for the Gaussian Orthogonal Ensemble of interacting states.   

If a system has integrals of motion, which takes place for our system of interest (see Sec. \ref{sec:Model})  the states with different values of the related integrals do not repel each other that would lead to the inevitable deviation from the Wigner-Dyson statistics even in the delocalized regime. To avoid this problem the states should be split into subgroups with a certain value of all integrals of motion \cite{ab17Comment}. For the periodic system and even number of atoms $N$ one can introduce $N+2$ such subgroups characterized by the quasi-momentum including $N-2$ subgroups with quasi-momenta different from $0$ or $N/2$ ($Q=-N/2+1,..-1,1,.. N/2-1$) and four subgroups with $Q=0, N/2$ either symmetric or anti-symmetric with respect to the reflection transformation. For an odd $N$ one has $N+1$ subgroups including $N-1$ subgroups with quasi-momenta ($Q=-(N-1)/2, -(N-1)/2+1..-1,1,.. (N-1)/2$) and two subgroups with $Q=0$ either symmetric or anti-symmetric with respect to the reflection transformation.
For other boundary conditions one has only two subgroups either symmetric or anti-symmetric with respect to the center of the chain. Since the results for symmetric and anti-symmetric states are quite similar, the level statistics reported below are related to the symmetric states. 

In contrast to spin or particle systems \cite{OganesyanHuse07,ab15MBL} we cannot directly apply exact diagonalization method to Hamiltonians in Eq. (\ref{eq:H}) because they have infinite basis of states since the vibration population numbers can take infinite number of values. Therefore, the basis states should be restricted to the finite number of states as described in the following section. 

\subsection{Basic Approximation} 
\label{sec:approx}

For any boundary conditions and specific subgroup of states the Hamiltonian in Eq. (\ref{eq:H}) can not be exactly diagonalized since the total number of possible basis states is infinite. To avoid this complexity the off-diagonal anharmonic interaction is restricted to the terms conserving the total number of excited quanta, $n_{t}$, similarly To Ref. \cite{Cohen13}. This means that only terms having two $\widehat{b}^{\dagger}$ and two $\widehat{b}$ operators are taken into consideration. Similar terms are left for other boundary conditions. 
This approximation should be valid at least qualitatively if the annahrmonic interaction  is weak. Consequently,  the anharmonic interaction energy $Bn_{t}^2/N$ should  be less than the harmonic interaction energy $n_{t}$ that yields 
\begin{eqnarray}
Bn_{t} <1. 
\label{eq:anh-val}
\end{eqnarray}

\begin{figure}[H]
\centering
\includegraphics[width=8 cm]{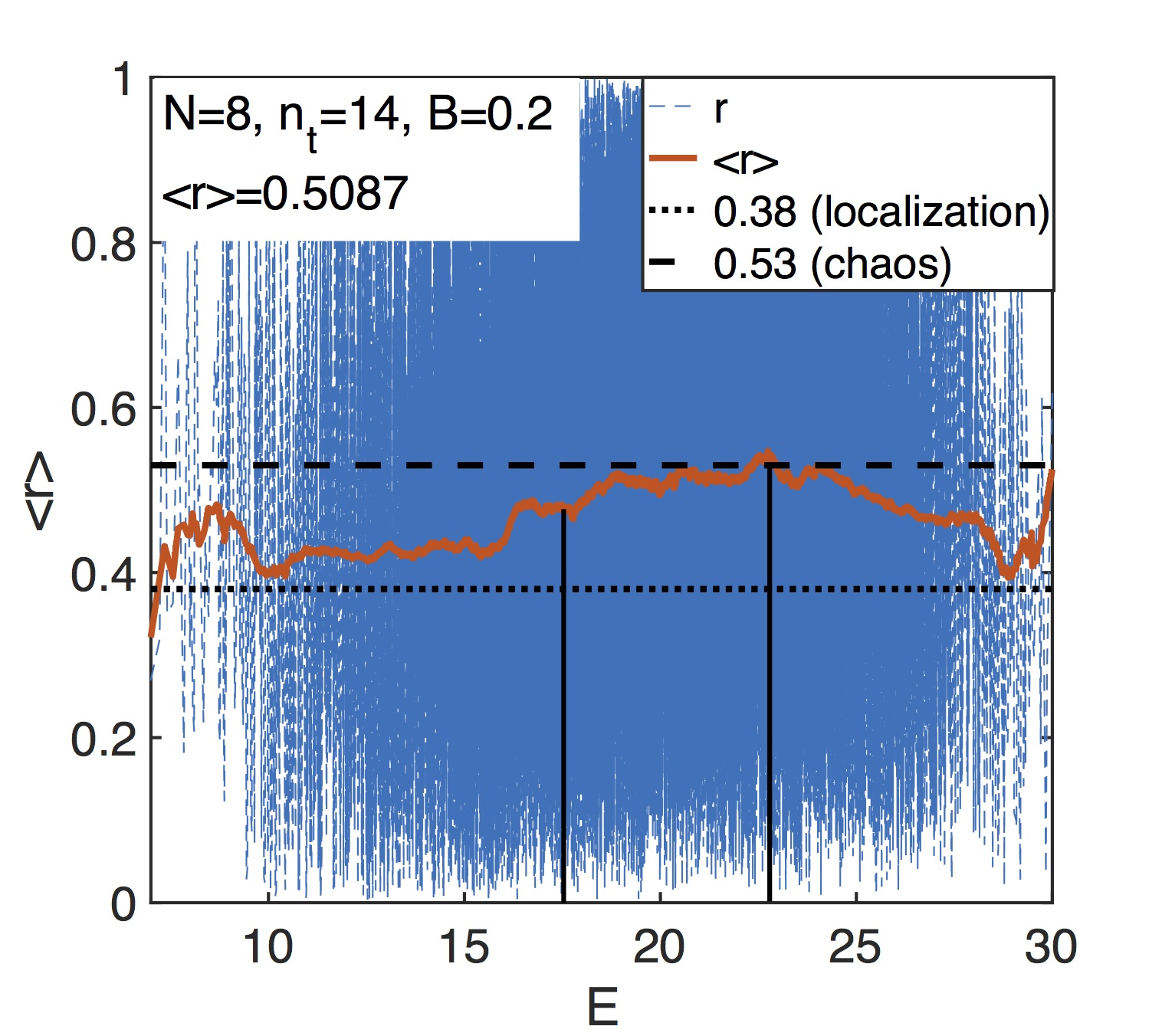}
\caption{Level statistics represented by the minimum ratio $r$, Eq. (\ref{eq:Og}) including $r$ as it is and  average ratio $<r>$ over $972$ adjacent eigenstates that clearly tends to the chaotic behavior $<r> \approx 0.53$ at the maximum density of states. The average level statistics $<r>=0.5087$ for the chain containing $8$ atoms, states with the number of phonons $n_{t}=14$ and anharmonicity strength $B=0.2$ averaged for mean $1/2$ of eigenstates located between two vertical lines.}
\label{fig:DataPoint}
\end{figure}

The modified Hamiltonian has a finite basis set for each specific number of atoms $N$ and number of phonons $n_{t}$ so it can be studied using the full diagonalization of the problem. The representative level statistics for the chain of $N$ atoms with free ends boundary conditions, total number of phonons $n_{t}=14$ and the strength of anharmonic interaction $B=0.2$ is shown in Fig. \ref{fig:DataPoint}. In contrast to the problems of particles or spins placed in a random potential \cite{OganesyanHuse07} where the ratio parameter $r$ can be averaged over many disorder realizations, here we have the only one realization of the system. In this realization  the ratio $r$ itself represents quasi-random number ranging between $0$ and $1$ in a chaotic manner as shown by the dashed dark blue line in  Fig. \ref{fig:DataPoint}. However, averaging the data over $972$ adjacent states ($5\%$ of the total number of states) leads to the smooth curve clearly approaching chaotic limit of $0.53$ near the middle of the spectrum. The average ratio $<r>$ for the given set of parameters $N, n_{t}, B$ has been determined   taking the arithmetic average of this minimum ratio over the middle half of the system eigenstates as shown in Fig. \ref{fig:DataPoint}.
This procedure describes how the data were collected to analyze the transition between the localized and chaotic regimes as a function of the number of atoms and phonons and the strength of anharmonic interactions. In other calculations the same averaging of the rato parameter $r$  is performed. 

The typical harmonic energy can be estimated for the given number $n_{t}$ of phonons using their sinusoidal dispersion law, Eq. (\ref{eq:Freq}), as $E_{h}=2n_{t}<|sin(x)|>=4n_{t}/\pi$. In the case of Fig. \ref{fig:DataPoint} this energy can be estimated as $17.83$. This energy is smaller than the typical average energy by around $10\%$ because of the anharmonic correction to the energy, which is still small.

The chosen representative states having maximum density  at fixed number of phonons do not perfectly represent the true thermodynamic states of the system at the given energy. In the classical regime $E>N$ (see Eq. (\ref{eq:maxN})) the thermodynamic average number of phonons scales as $E\ln(N)$ due to the contribution of low frequency phonons. We believe that this difference is not crucial since the logarithmic factor is related to low frequency phonons, which have substantially reduced anharmonic interaction strength and therefore can be ignored in the consideration of resonant interactions as discussed in Sec. \ref{sec:Reson}. The other reason is that the investigated states coexists with the ``thermodynamic" states at the same energy. If the states under consideration are chaotic, the other states at the same energy should be usually chaotic as well \cite{Huveneers16HotSpots}. 

In the quantum regime, $E<N$, the representation of the typical configuration by $E$ phonons with typical energy of order of unity is much less relevant than for the classical regime since the typical phonon energy is given by the thermal energy $\sqrt{E/N}$ that is much less than $1$. However, since the delocalization criterion, Eq. (\ref{eq:ResBelowDe}), is universal and does not depend on the number of phonons  we also believe that the theory should be applicable to the whole system at least quantitatively. 

Thus the numerical results reported below are preliminary and need improvement that  is postponed for the future.

The validity of the approach has been checked for the classical regime extending the basis to all states with the number of phonons less or equal to $n_{t}$. The results for this extension are consistent with those for the phonon number just equal to $n_{t}$. However, the calculations are much faster in the latter case and they permit us to obtain more conclusive results. The approach that seems to be more ``natural" restricting the basis to the states with energies less than a certain maximum energy $E_{max}$ works much worse and requires $E_{max} \sim 2E$ to give a reasonable estimate for the level statistics at energy $E$, which substantially limits our abilities to obtain conclusive results. This could be the consequence of broken connections due to the exclusion of significant states naturally present in the theory conserving the number of phonons.

\subsection{Dependence of localization transition on the boundary conditions and  the numbers of phonons and atoms.}
\label{sec:scaling}

Since our results below for the level statistics ($<r>$, see Eq. (\ref{eq:Og})) are expressed as a function of anharmonic interaction strength $B$ (see Fig. \ref{fig:NoChaos}, remember that only $\beta$ FPU problem is considered numerically), it is convenient to reexpress the criterion in Eq. (\ref{eq:crossResEarlier}) in terms of the critical strength $B_{c}$ dependence on the number of atoms $N$ and phonons $n_{t}$. Using Eq. (\ref{eq:maxN}) for the classical and quantum regimes we get 
\begin{eqnarray}
B_{c}=
\begin{cases}
  \frac{\pi^3}{2Nn_{t}}, & N < n_{t}, \\    
  \frac{\pi^3}{2n_{t}^2}, &    N > n_{t}.
\end{cases}
\label{eq:Bc}
\end{eqnarray}

\subsubsection{Effect of boundary conditions}

\begin{figure}[H]
\centering
\includegraphics[width=10 cm]{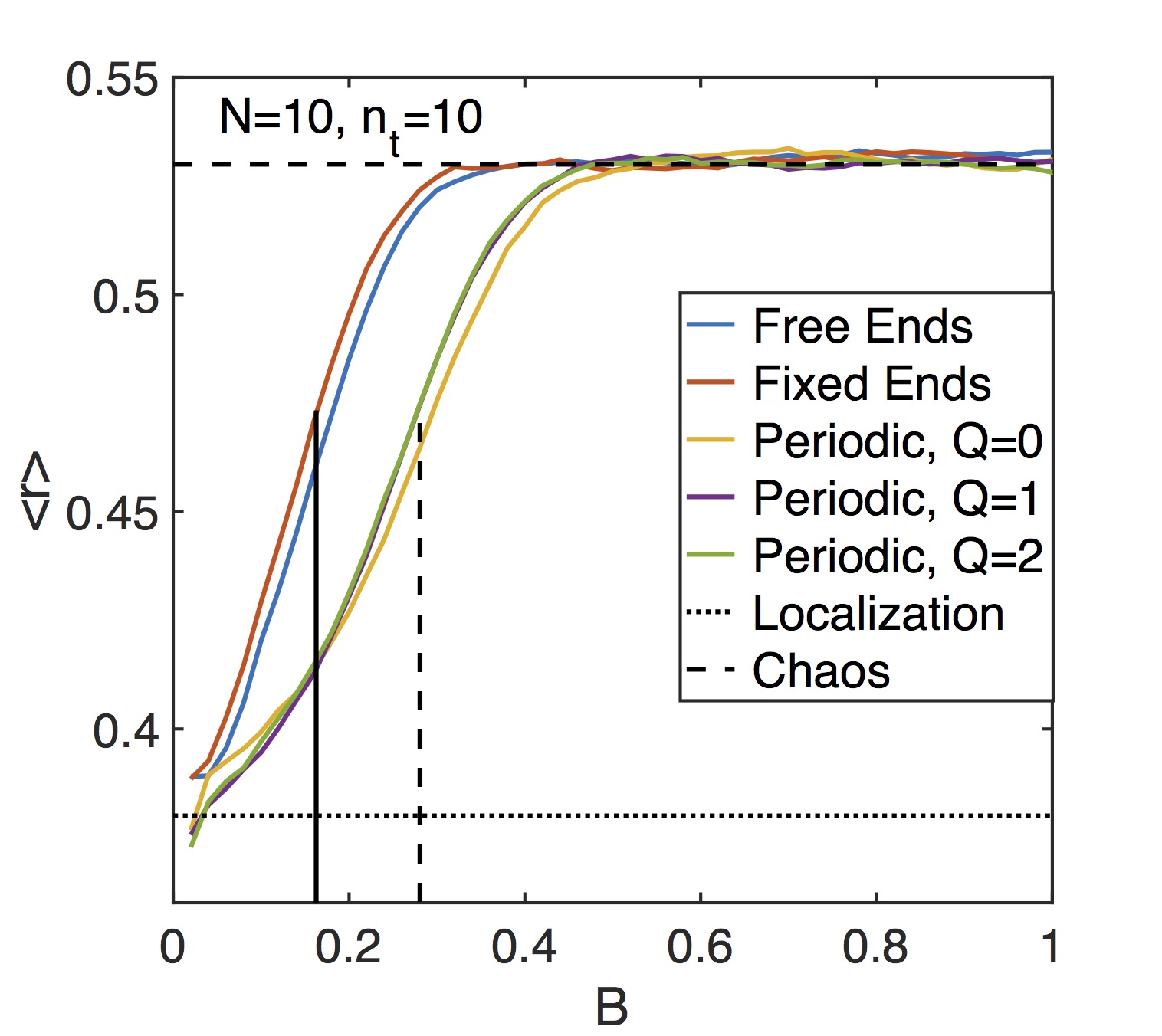}
\caption{Level statistics represented by the average minimum ratio $<r>$ for different boundary conditions and quasi-momenta. The transition point, $B_{c} \approx 0.15$ predicted by Eq. (\ref{eq:Bc}) is shown by the solid vertical line, while the dashed vertical line shows the transition point estimate for the periodic regime, $B_{c} \approx 0.28$.}
\label{fig:LStComp}
\end{figure}

To examine the effect of boundary conditions  consider some representative data obtained for the level statistics parameter $<r>$ vs. the strength of anharmonic interaction $B$ following the technique described in Sec. \ref{sec:approx} (see Fig. \ref{fig:DataPoint})  for the chain of  $N=10$ atoms with all possible boundary conditions and for quasi-momenta $Q=0$, $1$ and $2$ in the case of periodic conditions. These dependencies are shown in Fig. \ref{fig:LStComp}.


According to Fig. \ref{fig:LStComp} it is clear that at large anharmonicity $B>0.4$ the system  is chaotic, while it is integrable at small anharmonicity $B<0.15$. For $N=10$ and $n_{c}=10$ the criterion of Eq. (\ref{eq:Bc}) predicts $B_{c} \approx 0.15$. This number is in an excellent agreement with  the numerical results for free ends or fixed ends boundary conditions as indicated by the vertical line in Fig. \ref{fig:LStComp}. Assuming that  $B_{c} \approx 0.15$ characterizes the transition in the case of fixed or free ends boundary conditions one can estimate the threshold anharmonicity in the periodic system using similar criterion (see dashed vertical line in Fig. \ref{fig:LStComp}) as $B_{c,per} \sim 0.28$. This estimates the approximate difference between two critical anharmonicities as $B_{c,per}/B_{c} \approx 1.87$  and we will use this ratio for quantitative estimates of transition parameters below.  Probably,  theoretical analysis of Refs. \cite{BermanKolovskii84FPU,Dauxois97FPU} is less applicable to the periodic system because of additional integrals of motion there. The emergence of chaotic phase in periodic system at larger anharmonicity compared to other boundary conditions can be the consequence of the smaller effective phase space in the former case due to additional integrals of motion lacking in the latter case. On the other hand, the results for periodic conditions are almost insensitive to the quasi-momentum, and the results for fixed and free ends boundary conditions are also quite similar to each other. Therefore in our predictions for localization threshold we do not distinguish between the free and fixed end boundary conditions as well as between different quasi-momenta for periodic boundary conditions. 

However, it is necessary to distinguish between periodic boundary conditions and others. We suggest the simplest form of difference redefining Eq. (\ref{eq:Bc}) reasonably valid for fixed and free ends regimes by multiplying its left hand side by the factor $0.28/0.15$ in agreement with the numerical results in Fig. \ref{fig:LStComp}.  Then for the periodic system Eq. (\ref{eq:Bc}) should be modified as 
\begin{eqnarray}
B_{c,per}=
\begin{cases}
  \frac{0.93\pi^3}{Nn_{t}}, & N < n_{t}, \\    
  \frac{0.93\pi^3}{2n_{t}^2}, &    N > n_{t}.
\end{cases}
\label{eq:Bcper}
\end{eqnarray}
Consequently, one should modify the critical energy behavior predicted by Eq. (\ref{eq:ANSGen}) as 
\begin{eqnarray}
E_{c,per}=
\begin{cases}
  \frac{1.07\pi^2}{N\left(B-\frac{4A^2}{9}\right)}, & N < \frac{1.035 \pi}{\sqrt{B-\frac{4A^2}{9}}}, \\    
  \frac{1.035 \pi}{\sqrt{B-\frac{4A^2}{9}}}, &    N > \frac{1.035 \pi}{\sqrt{B-\frac{4A^2}{9}}}.
\end{cases}
\label{eq:ANSGenPer}
\end{eqnarray}

\subsubsection{Dependence of localization threshold on numbers of atoms and phonons.}

Consider the dependence of the threshold anharmonicity on the energy expressed through the number of phonons. Most of the data are presented for periodic chains because the large number of integrals of motion there  reduces the total number of states permitting us to investigate larger numbers of atoms and phonons compared to other boundary conditions. 

For a demonstration of the method we consider periodic chain for $N=10$ atoms with possible numbers of phonons $n_{t}=9$, $10$, $11$ and $12$. The subgroup of symmetric states with quasi-momentum $Q=0$ is considered. 

\begin{figure}[h!]
\centering
\subfloat[]{\includegraphics[scale=0.12]{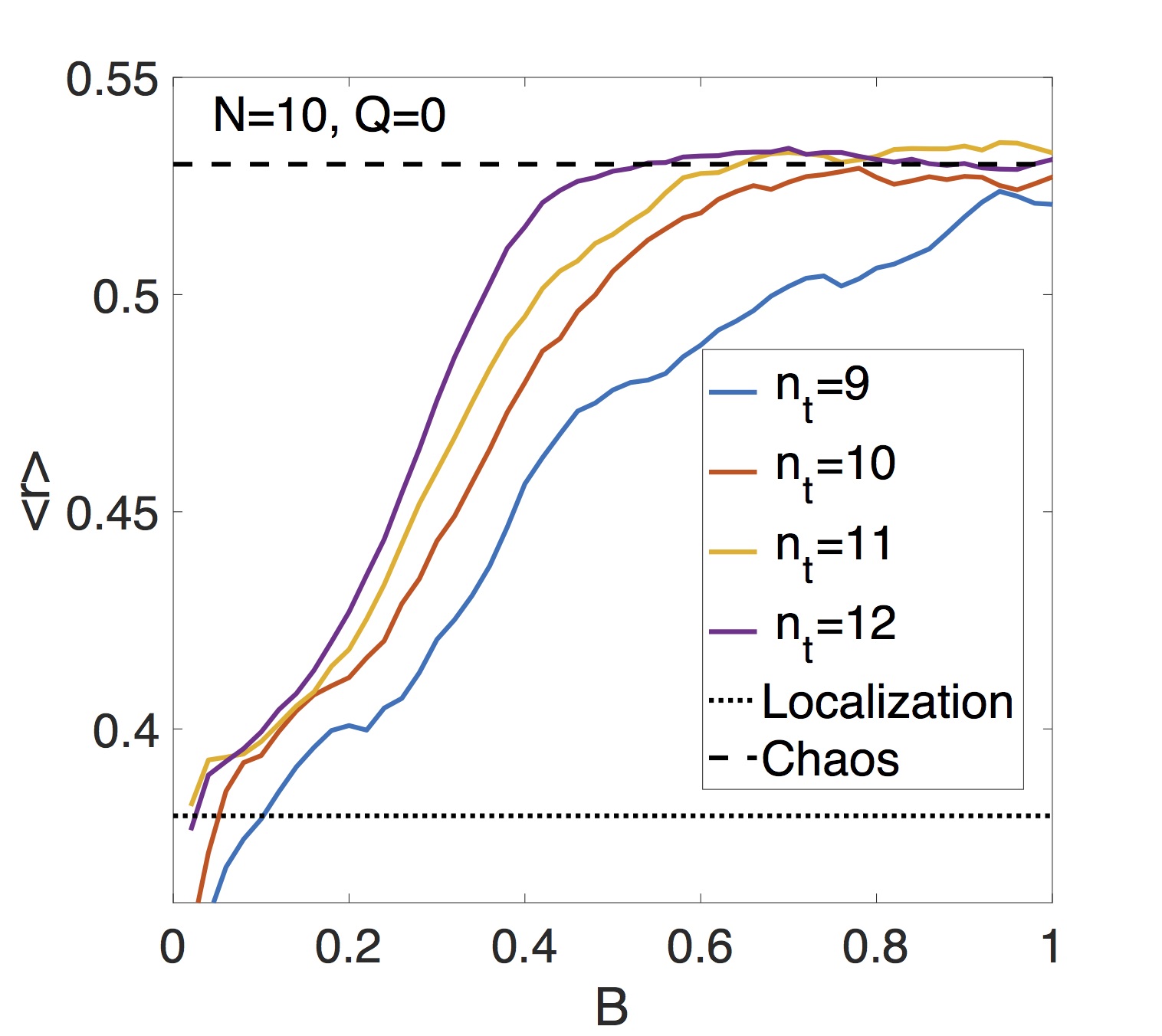}}
\subfloat[]{\includegraphics[scale=0.12]{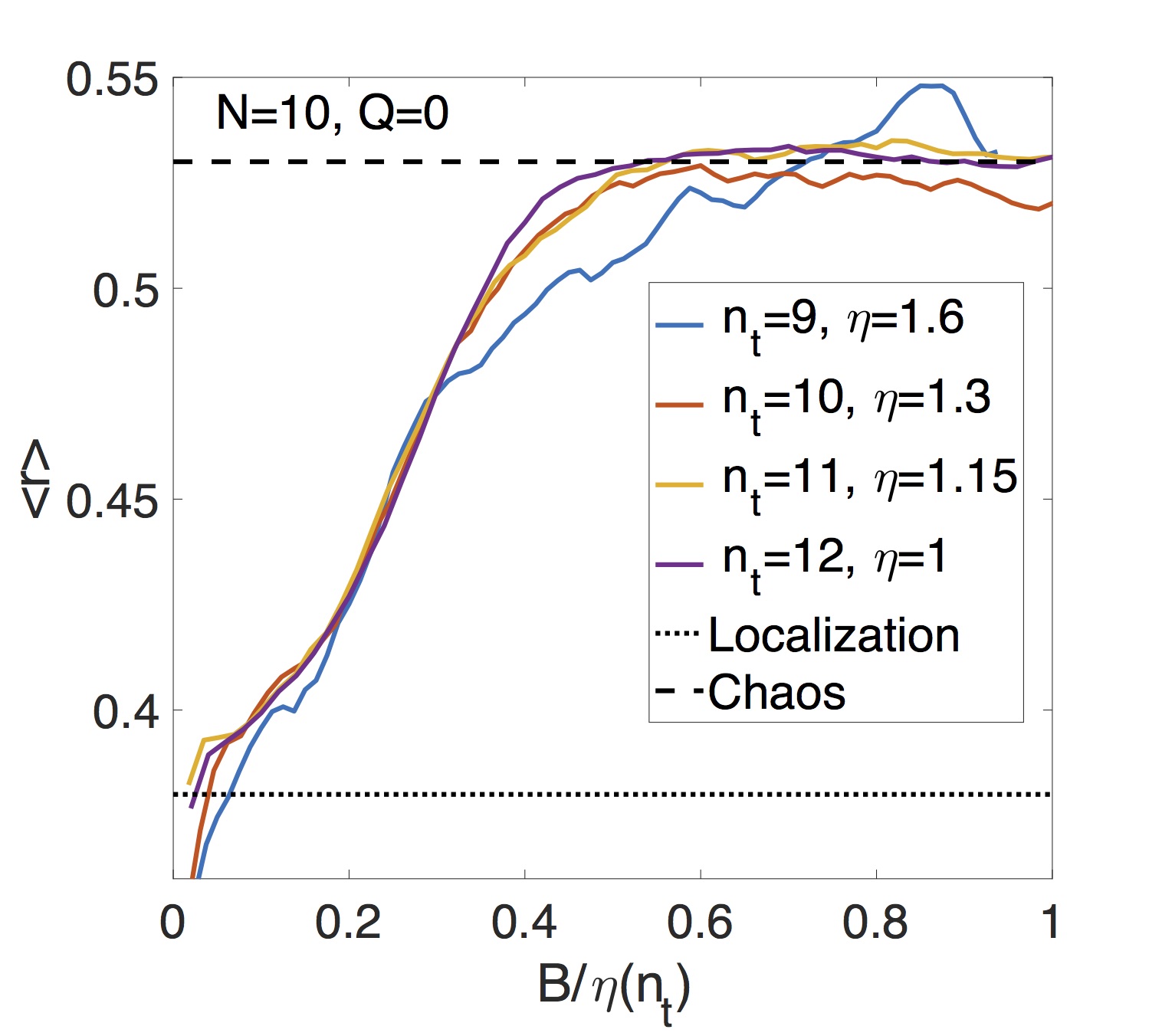}}
{\caption{Level statistics dependence on the anharomonic interaction for periodic chain of $N$ atoms and different total number of phonons as it is (a) or rescaled to attain the optimum match between the data (b).}
\label{fig:NoChaos}}
\end{figure}

The choice of possible parameters $n_{t}$ is limited because of the poor data averaging for small size of phase space less than $5000$ states and exponential increase of the number of states with increasing $n_{t}$. Indeed, for $n_{t}=9$ the basis contains $4420$ states that is insufficient for good averaging of level statistics as it is seen from Fig. \ref{fig:NoChaos}. a, while for $n_{t}=12$ the basis contains $26720$ states that is closed to the maximum matrix size where exact diagonalization can still be performed using standard MATLAB algorithms.


\begin{figure}[H]
\centering
\includegraphics[width=10 cm]{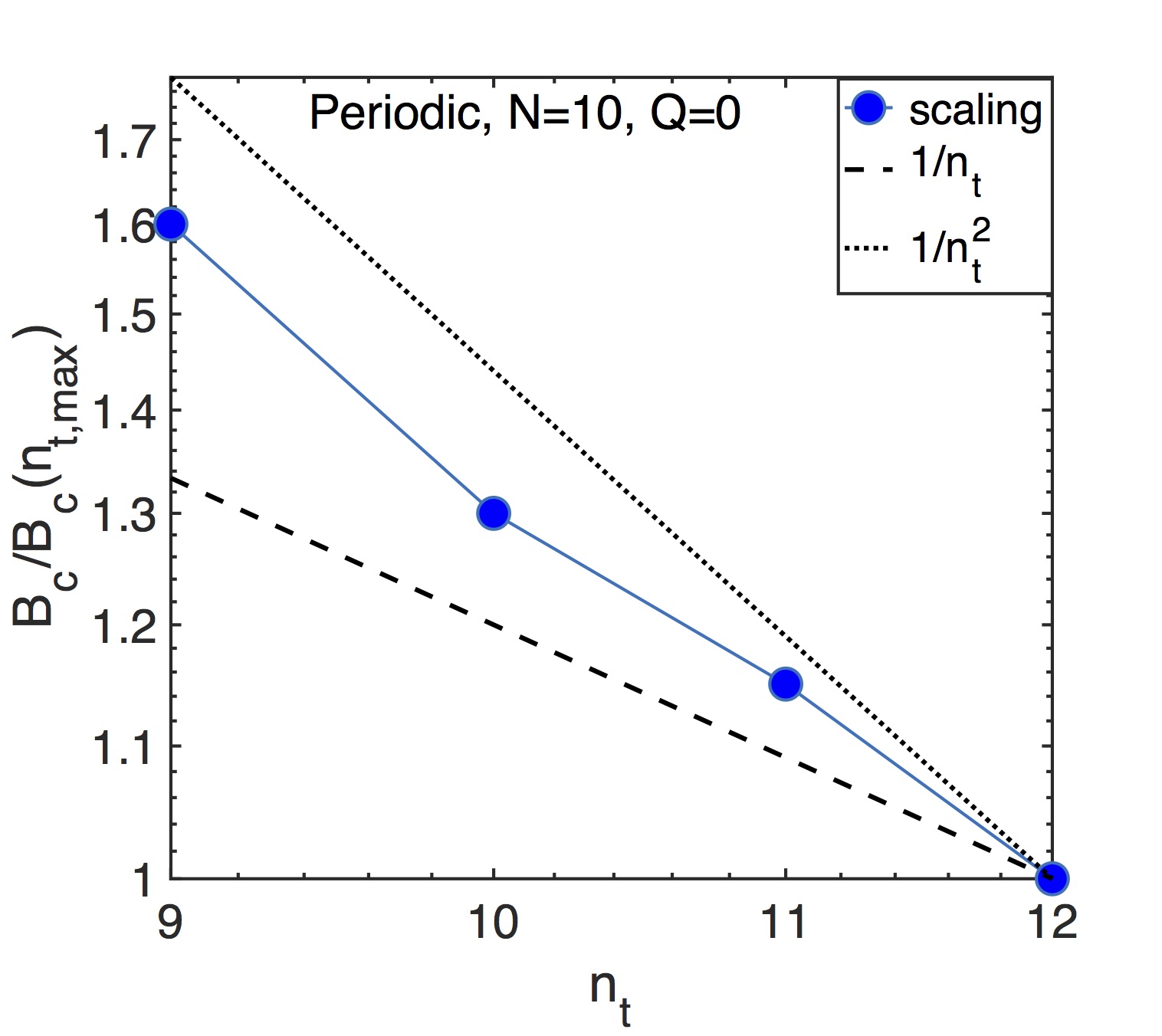}
\caption{Scaling of critical anharmonicity with the number of phonons (circles) as compared with the theory predictions in classical and quantum regimes, Eq. (\ref{eq:Bc}),  for $N=10$ atoms.}
\label{fig:ntScalingN10Q0}
\end{figure}

 To determine the algebraic dependence of localization threshold on the number of phonons  we use the data rescaling procedure similarly to the earlier work in spin systems \cite{ab15MBL,ab15MBLXY,ab17Comment}. This procedure attempts  to attain the maximum match between different data rescaling the $x$ axis. As it is shown in Fig. \ref{fig:NoChaos}.b, the reasonable match can be attained rescaling the data for different $n_{t}$ with respect to those for maximum $N=12$ by the $n_{t}$-dependent parameter $\eta$ shown in Fig. \ref{fig:NoChaos}.b. The scaling of parameter $\eta(n_{t})$ is related to that of a critical anharmonicity $B_{c}(n_{t})$  as 
\begin{eqnarray}
B_{c}(n_{t})=B_{c}(n_{t,max})\eta(n_{t}).
\label{eq:ScBcnt}
\end{eqnarray}
In our case $n_{t,max}=12$. Consequently, we end up with the dependence of the critical anharmonicity $B_{c}$ on the number of phonons, $n_{t}$ as shown in Fig. \ref{fig:ntScalingN10Q0}.

\begin{figure}[h!]
\centering
\subfloat[]{\includegraphics[scale=0.12]{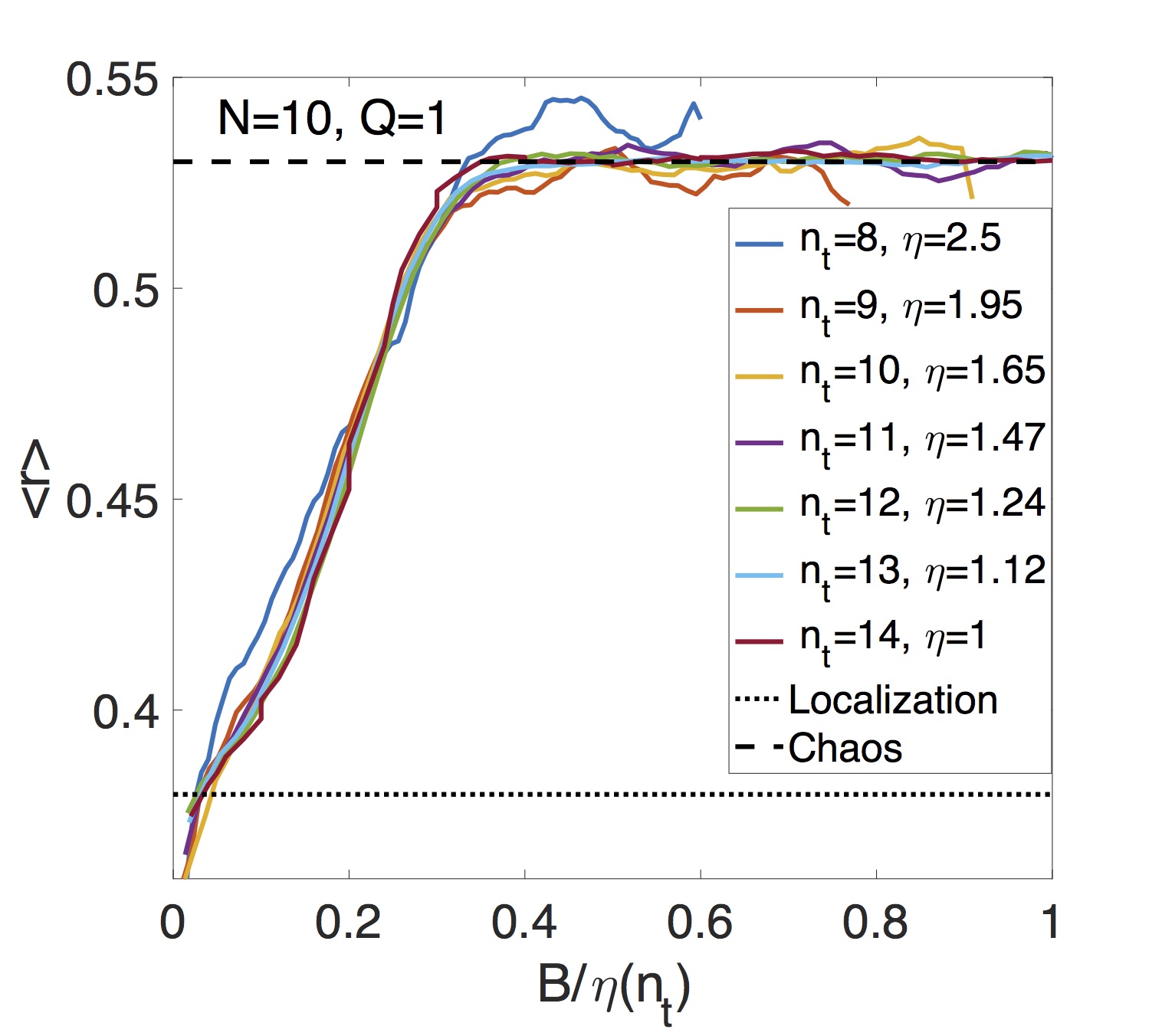}}
\subfloat[]{\includegraphics[scale=0.12]{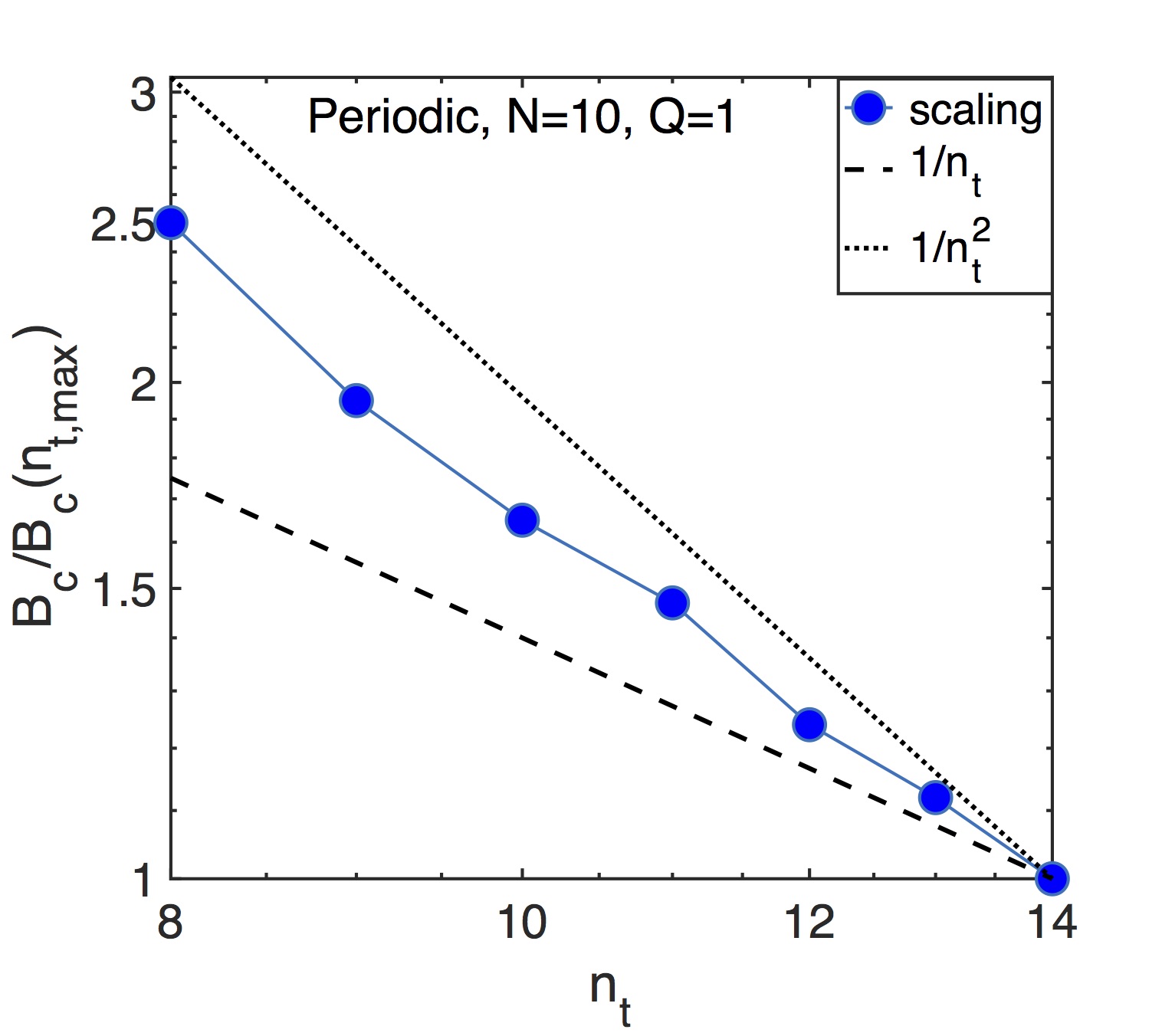}}
{\caption{(a) Match of level statistics at different numbers of phonons for the intermediate number of atoms $N=10$. (b) Scaling of critical anharmonicity with the number of phonons (circles)  for $N=10$ atoms.}
\label{fig:ntDepN10Q1}}
\end{figure}

The observed dependence is in between two predictions of Eq. (\ref{eq:Bc})  that is not surprising because the calculations are made for $n_{t} \sim N=10$ near the crossover between classical and quantum regimes. This justifies our definition of that crossover in Eqs. (\ref{eq:ANSGen}), (\ref{eq:ANSGenPer}).  Similar behavior takes place for the same number of atoms and quasi-momentum $Q=1$ as shown in Fig. \ref{fig:ntDepN10Q1}.

\begin{figure}[h!]
\centering
\subfloat[]{\includegraphics[scale=0.12]{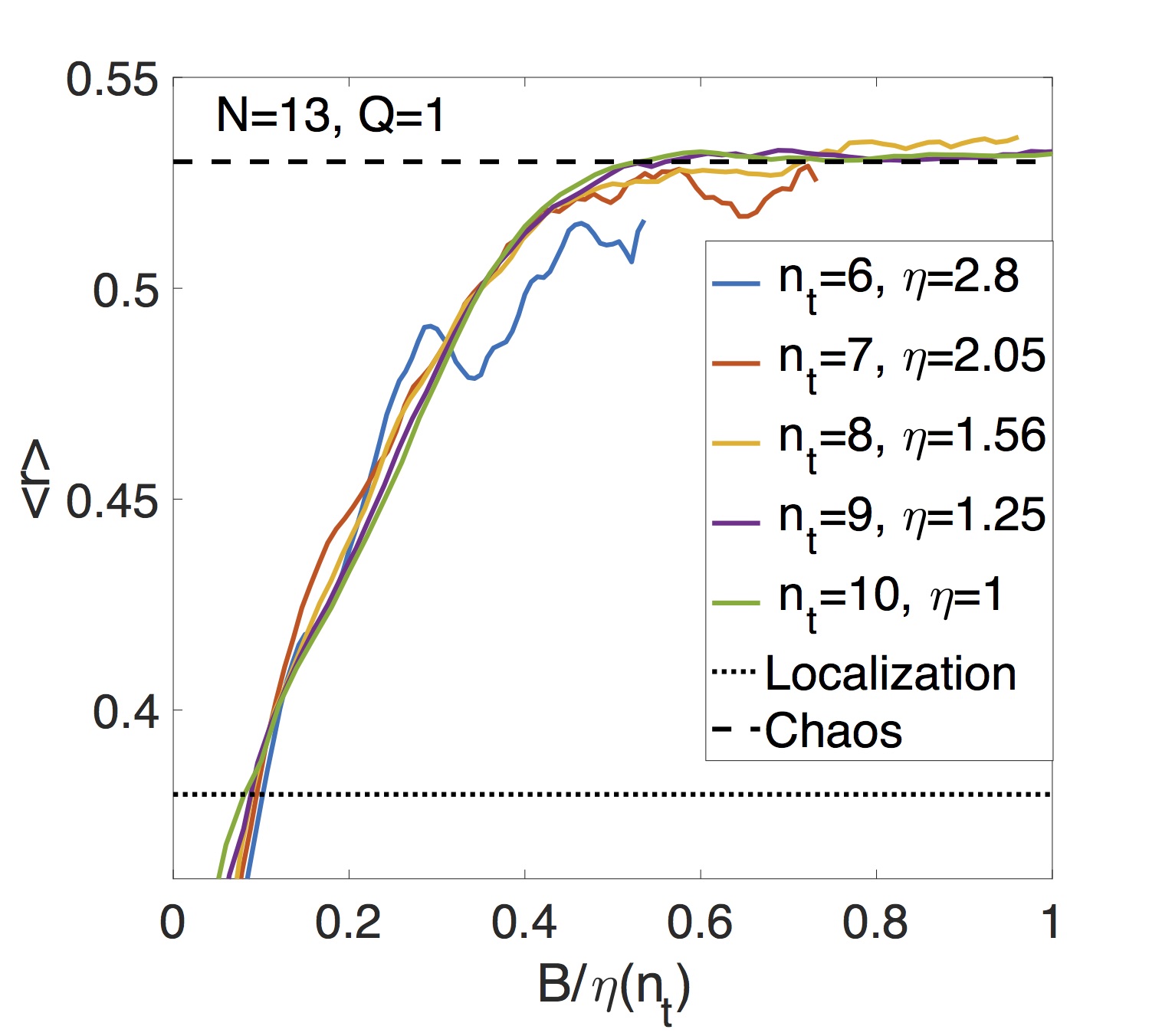}}
\subfloat[]{\includegraphics[scale=0.12]{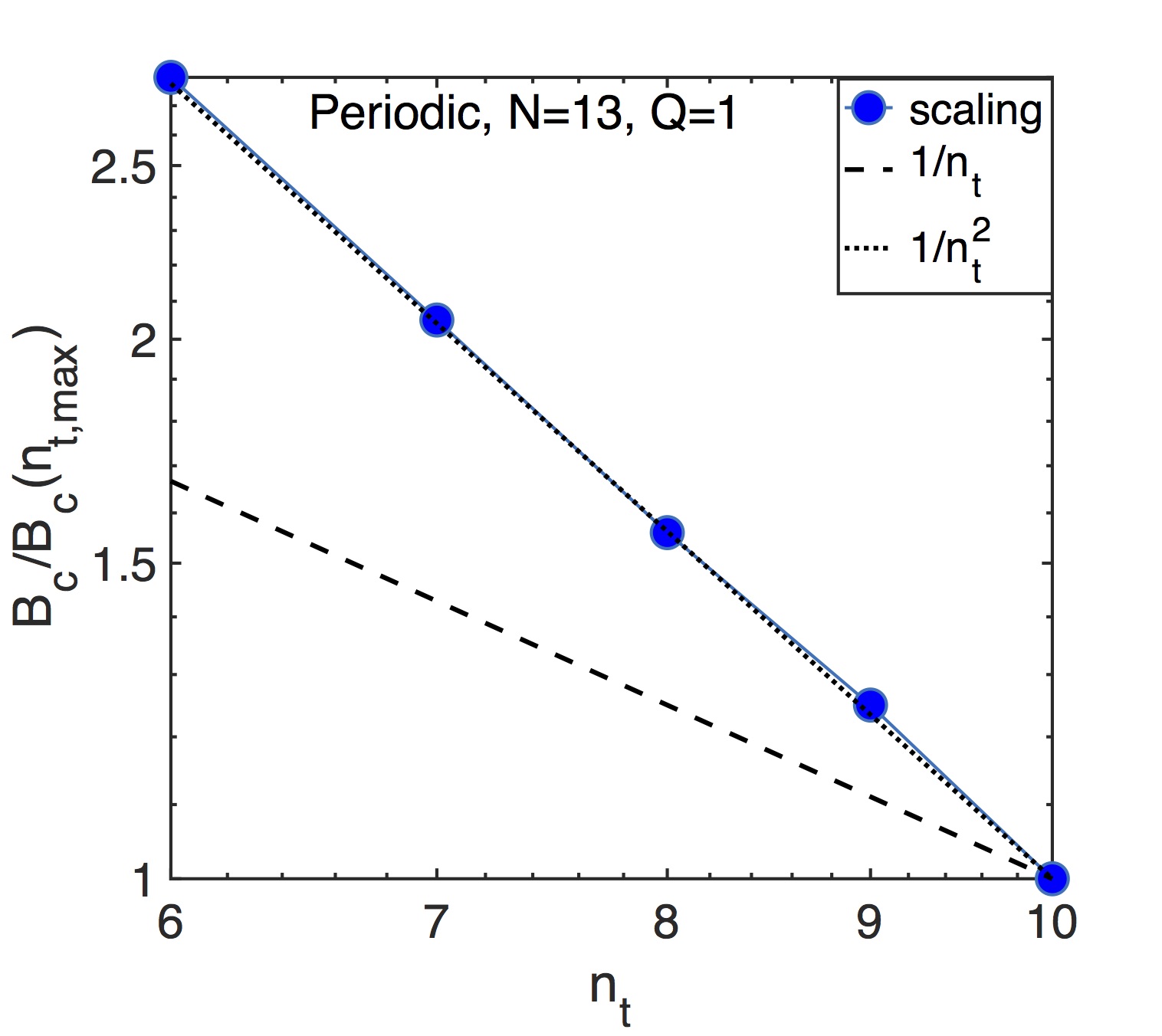}}
{\caption{(a) Match of level statistics at different numbers of phonons for the large number of atoms $N=13$. (b) Scaling of critical anharmonicity with the number of phonons (circles)  for $N=13$ atoms. (Periodic boundary conditions, $Q=1$.}
\label{fig:ntDepN13Q1}}
\end{figure}

Consider limiting quantum ($n_{t} <N$) and classical ($n_{t}> N$) regimes. Representative results for quantum regime are shown in  Fig. \ref{fig:ntDepN13Q1} in the case of $N=13$ atoms and a number of phonons, $n_{t}$, ranging from $6$ to $10$. The observations clearly agree with Eq. (\ref{eq:Bc})  for $n_{t}<N$. 

\begin{figure}[h!]
\centering
\subfloat[]{\includegraphics[scale=0.12]{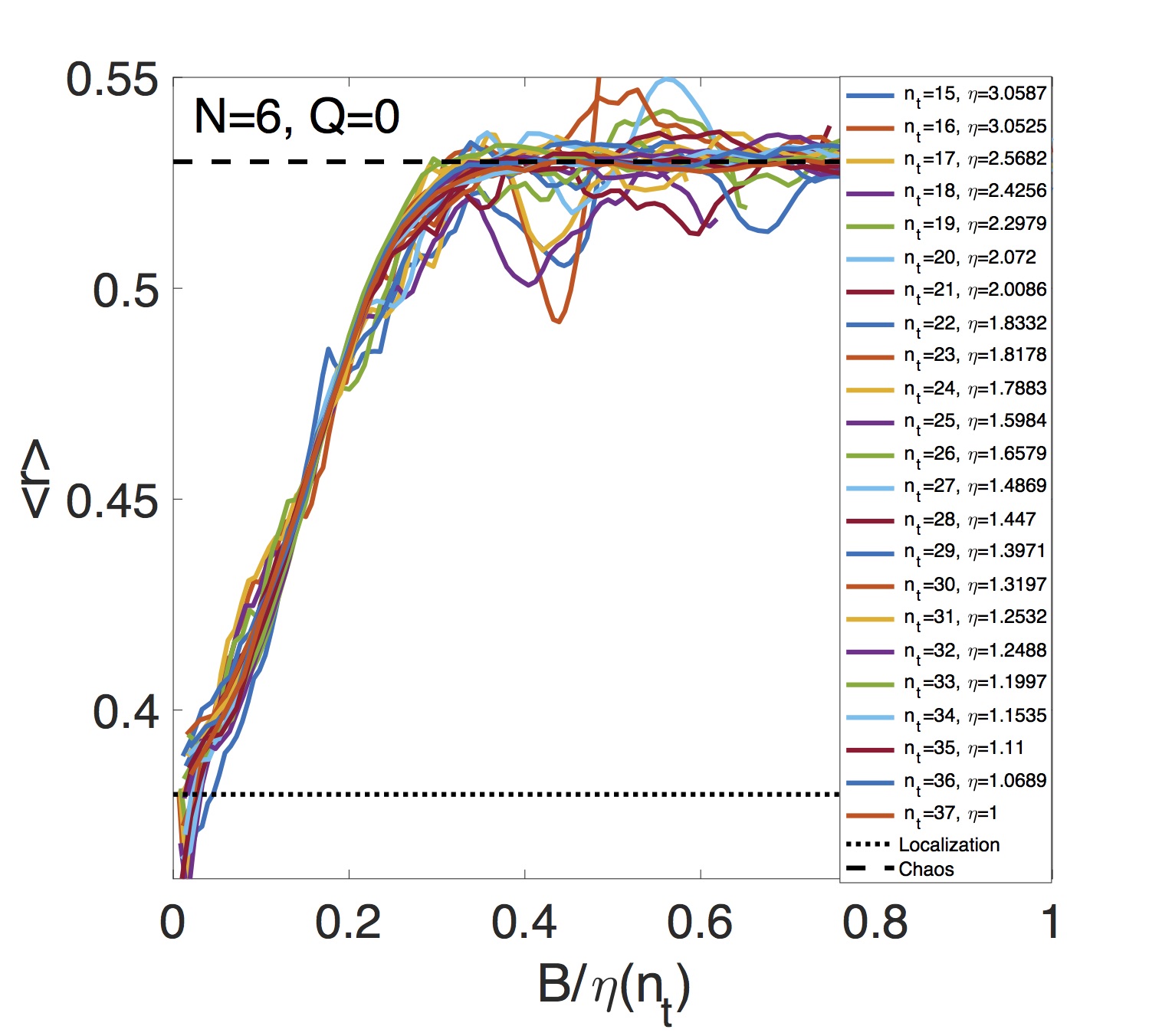}}
\subfloat[]{\includegraphics[scale=0.12]{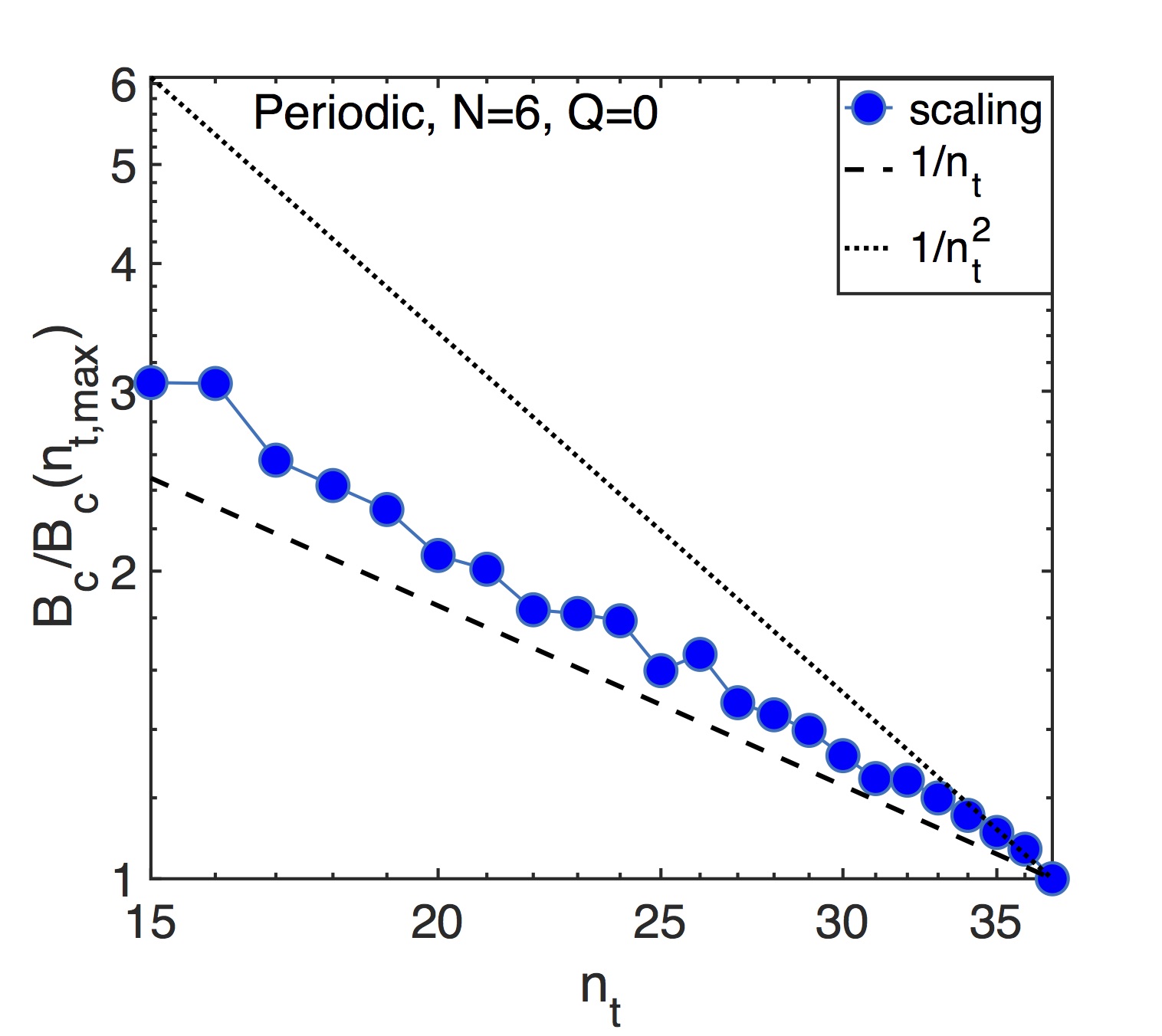}}
{\caption{(a) Match of level statistics at different numbers of phonons for the small number of atoms $N=6$. (b) Scaling of critical anharmonicity with the number of phonons (circles)  for $N=6$ atoms (periodic boundary conditions, $Q=0$).}
\label{fig:ntDepN6Q1}}
\end{figure}

The opposite, classical limit, $n_{t} > N$, is represented by the periodic chain of $N=6$ atoms considered for the number of phonons, $n_{t}$, ranging from $18$ to $36$. The obtained dependence shown in Fig. \ref{fig:ntDepN6Q1} is very close to the inverse proportionality, Eq. (\ref{eq:Bc}), valid in this limit. The growing deviation at small $n_{t}$ is probably caused by quantum effects, significant for $n_{t} \sim N$.

\begin{figure}[h!]
\centering
\subfloat[]{\includegraphics[scale=0.12]{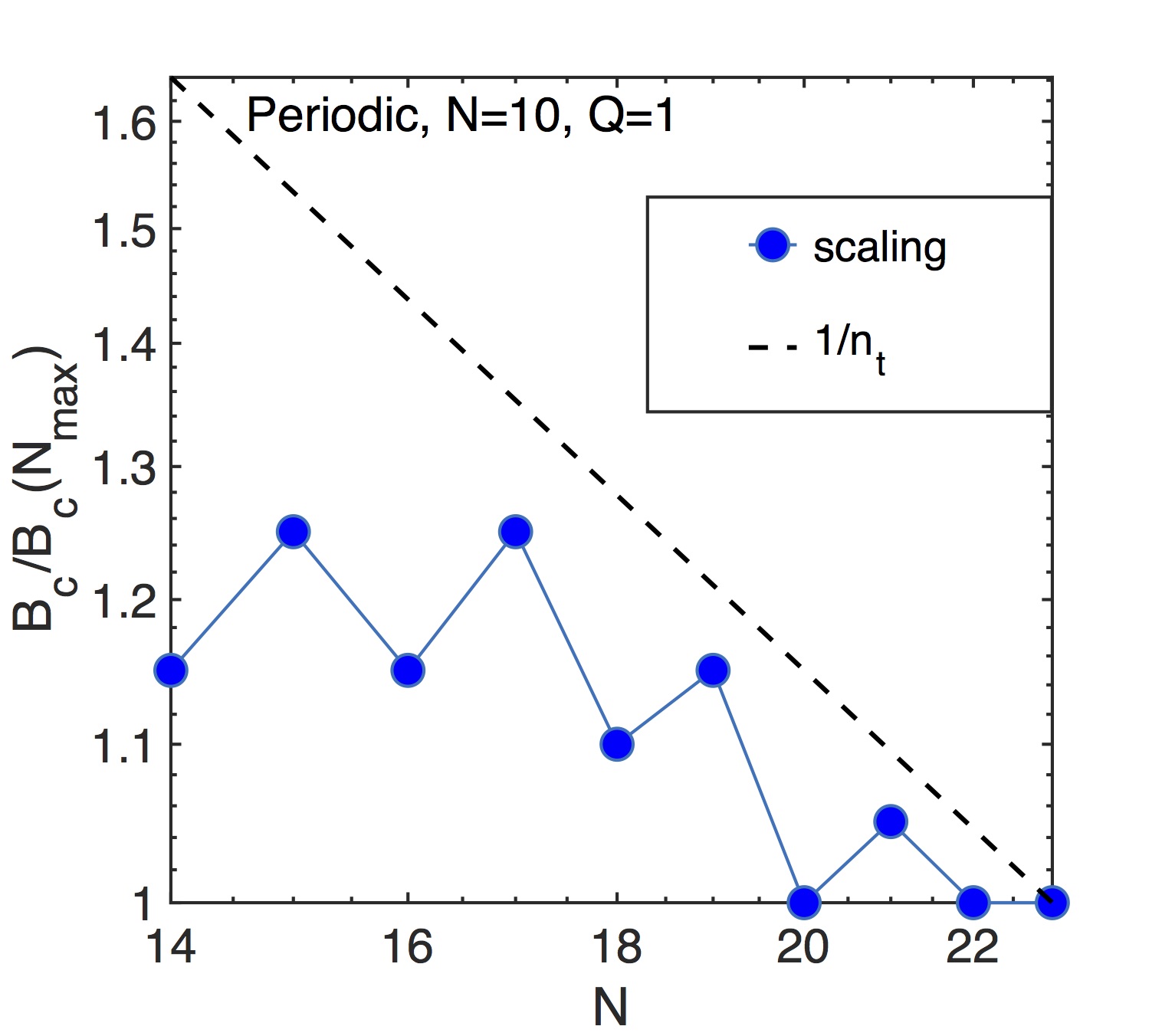}}
\subfloat[]{\includegraphics[scale=0.12]{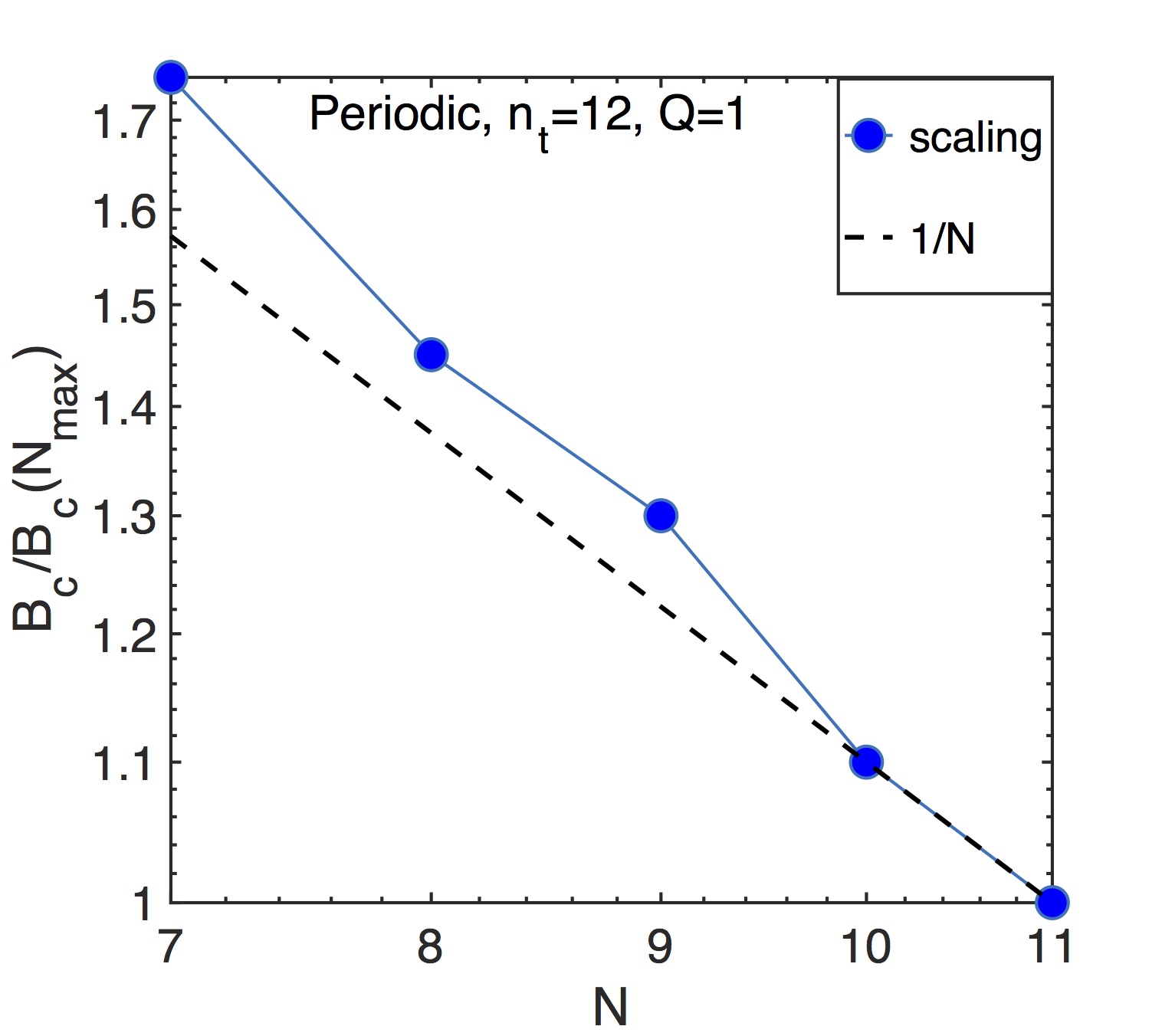}}
{\caption{Scaling of critical anharmonicity with the number of atoms  (circles) as compared with theory predictions in classical and quantum, Eq. (\ref{eq:Bc}), regimes for $n_{t}=6$ (a) and $n_{t}=12$ (b)  phonons.}
\label{fig:NScalingnt6Q1}}
\end{figure}

Similarly, one can consider the dependence of the threshold anharmonicity, $B_{c}$, on the number of atoms $N$ at fixed number of phonons $n_{t}$. This dependence is expected to be an inverse proportionality in the classical regime of a large number of phonons, $n_{t}>N$, while no dependence is expected in the opposite, quantum  limit of a small number, see Eq. (\ref{eq:Bc}). These expectations are consistent with the results given in Fig. \ref{fig:NScalingnt6Q1} in the quantum ($n_{t}=6$, Fig. \ref{fig:NScalingnt6Q1}.a) and classical ($n_{t}=12$, Fig. \ref{fig:NScalingnt6Q1}.b) limits.

\begin{figure}[h!]
\centering
\subfloat[]{\includegraphics[scale=0.12]{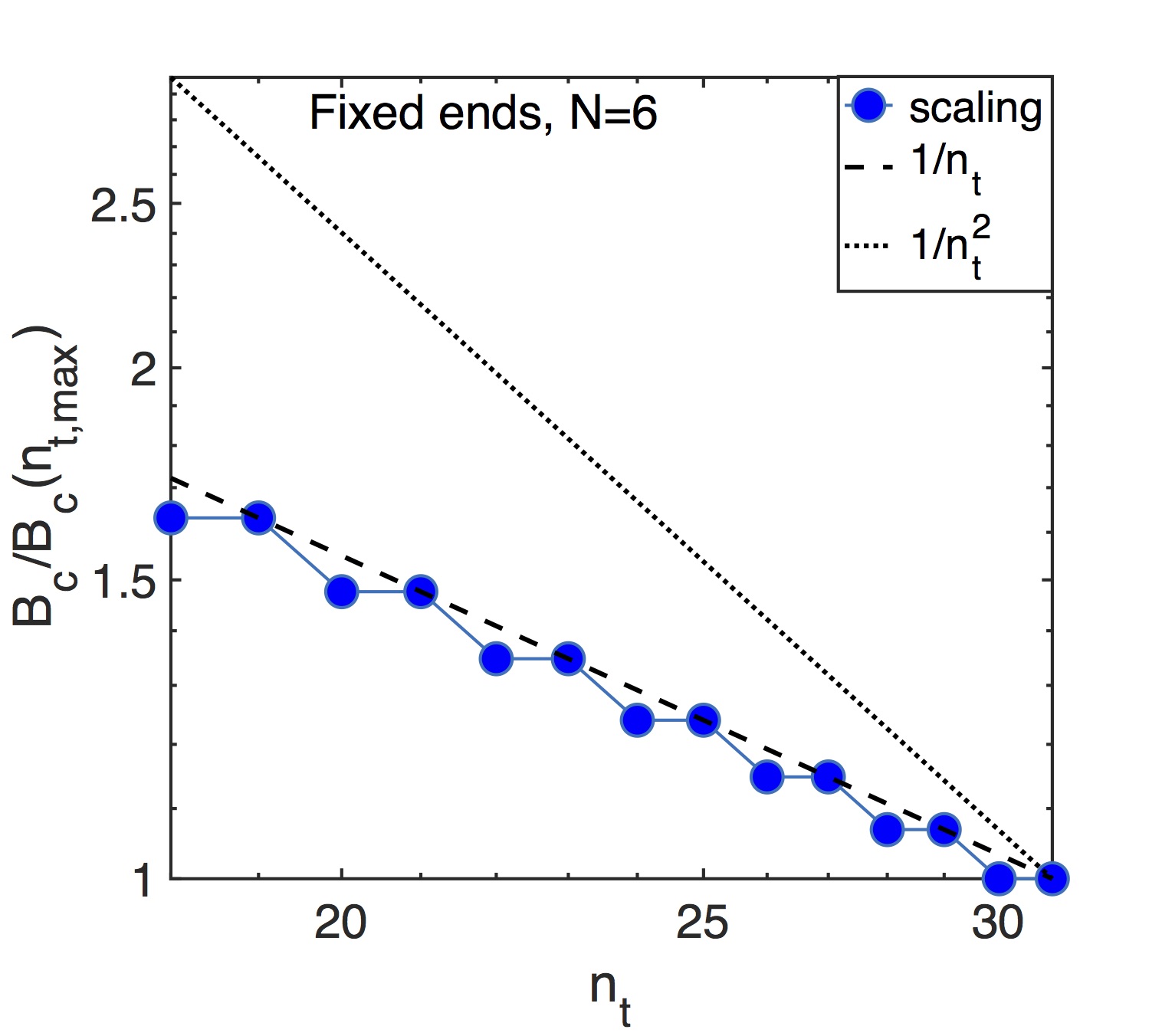}}
\subfloat[]{\includegraphics[scale=0.12]{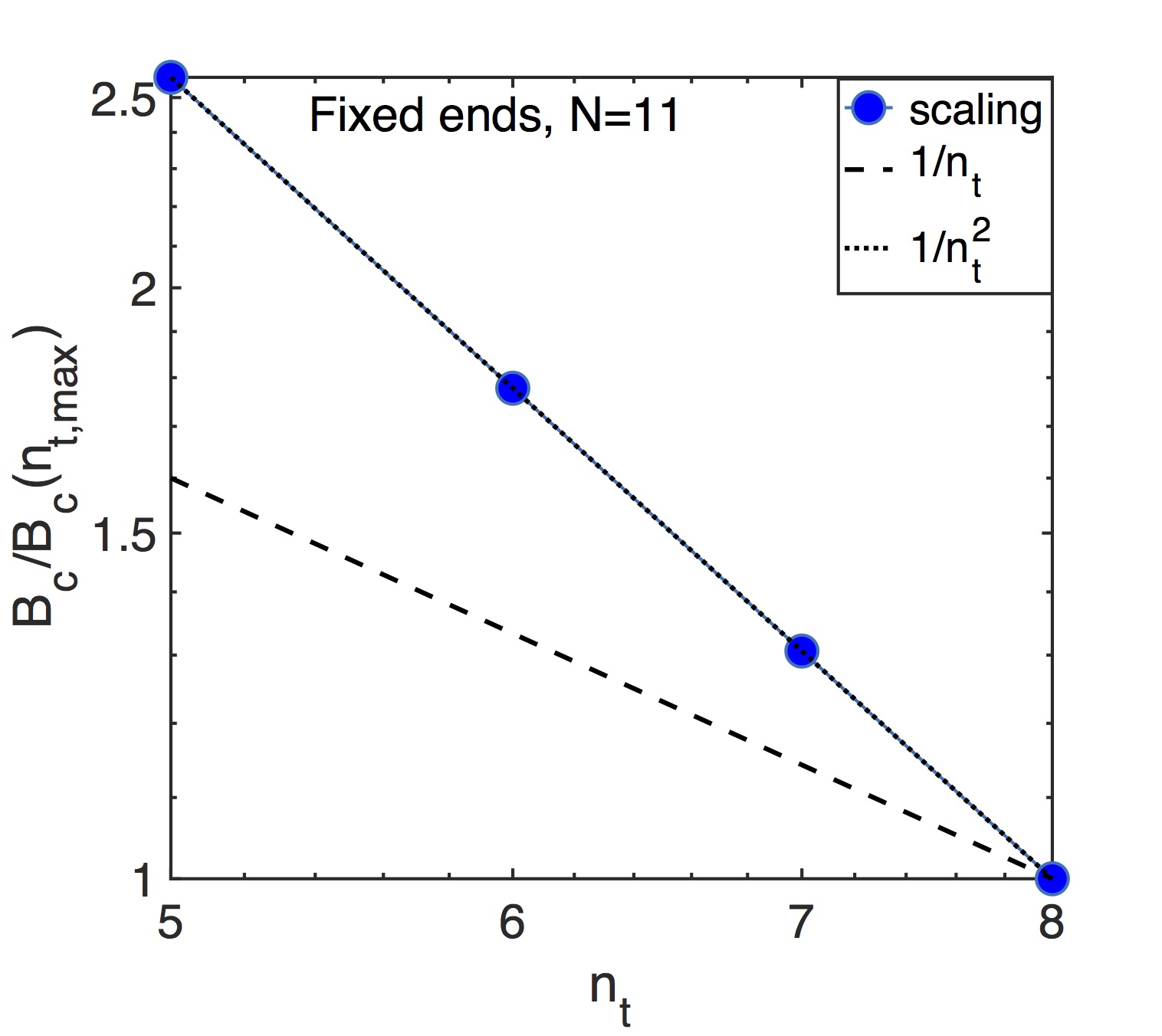}}
{\caption{Scaling of critical anharmonicity with the number of phonons  (circles) as compared with theory predictions in classical, Eq. (\ref{eq:Bc}), and quantum, Eq. (\ref{eq:EcSmallN}) regimes for $N=6$ (a) and $N=11$ (b)  atoms in the chain with fixed ends.}
\label{fig:NScalingFxE}}
\end{figure}

The results for other boundary conditions are also consistent with theoretical predictions as illustrated in Fig. \ref{fig:NScalingFxE} both in classical ($N=6$) and quantum ($N=11$) regimes.



Thus the numerical investigation of localization - chaos transition supports the theory predictions, Eqs. (\ref{eq:ANSGen}), (\ref{eq:ANSGenPer}).

\section{Discussion}
\label{sec:disc}

Here we reformulate the results in terms of standard notations in Table \ref{tab}, and attempt to apply them to organic molecules. We predicted the threshold energies for emergence of chaotic dynamics for combined $\alpha+\beta$ FPU problem  as a function of anharmonic interaction strengths and system sizes as given by Eqs. (\ref{eq:ANSGen}), (\ref{eq:ANS}). 

For practical application of these results it is convenient to reexpress them in terms of the dimensional force constant $k$, atomic mass $M$ and Planck constant $\hbar$. This requires the change of anharmonic interaction constants as $B \rightarrow B/k^2$, $A\rightarrow A^2/k^3$ in classical estimates and modify the critical energy as $N \rightarrow N\hbar \sqrt{k/M}$. The results are presented in Table \ref{tab} in the standard notations.

\begin{table}[H]
\caption{Summary of the results for localization threshold in classical and quantum regimes and definitions of those regimes in reduced and standard notations.}
\centering
\begin{tabular}{ccccc}
\toprule
\textbf{Model and Regime	}	 & \textbf{$\alpha+\beta$, classical}	& \textbf{$\alpha+\beta$, quantum}\\
\midrule
$E_{c}$, periodic	 & $\frac{3.73\pi^2 k^2}{N\left(B-\frac{4A^2}{9k}\right)}$			& $\frac{1.93\pi\hbar^{1/2}k^{5/4}}{\sqrt{B-\frac{4A^2}{9k}}M^{1/4}}$\\
Parametric domain  & $N<\frac{1.93\pi k^{3/4}M^{1/4}}{\hbar^{1/2}\sqrt{B-\frac{4A^2}{9k}}}$		& $N>\frac{1.93\pi k^{3/4}M^{1/4}}{\hbar^{1/2}\sqrt{B-\frac{4A^2}{9k}}}>1$	\\
$E_{c}$, free or fixed ends	 & $\frac{2\pi^2 k^2}{N\left(B-\frac{4A^2}{9k}\right)}$			& $\frac{1.41\pi\hbar^{1/2}k^{5/4}}{\sqrt{B-\frac{4A^2}{9k}}M^{1/4}}$\\
Parametric domain  & $N<\frac{1.41\pi k^{3/4}M^{1/4}}{\hbar^{1/2}\sqrt{B-\frac{4A^2}{9k}}}$			& $N>\frac{1.41\pi k^{3/4}M^{1/4}}{\hbar^{1/2}\sqrt{B-\frac{4A^2}{9k}}}>1$	\\
\bottomrule
\end{tabular}
\label{tab}
\end{table}

One can attempt  to apply these results to organic molecules using parameters for C$-$C bond extracted from the Morse potential \cite{MorseLinnett1941} that can be defined in terms of bond dissociation  energy $E_{d}=5.78\cdot 10^{-19}$J and inverse interaction radius $\alpha=3.45\cdot 10^{10}$m$^{-1}$ as 
\begin{eqnarray}
k=2E_{d}\alpha^2, ~ A=-6E_{d}\alpha^3, ~ B=14E_{d}\alpha^4. 
\label{eq:MorsePars}
\end{eqnarray}
Consequently, the expressions for the threshold energy in classical and quantum regimes for either free or fixed ends boundary conditions can be written as  $E_{c,cl}=4\pi^2 E_{d}/(3N)$ and $E_{c,q}=2\pi\sqrt{3}\sqrt{E_{d}\hbar\sqrt{k/M}}$, where $M$ is the atomic mass. The transition between two regimes takes place at the number of atoms $N_{c} \approx 2\pi\sqrt{3}\sqrt{E_{d}/(\hbar\sqrt{k/M})}$. The chaos can take place in the stable molecular state at energy less than the dissociation energy that is true only for  sufficiently long molecules containing $N\geq 14$ atoms. This is the result for atomic interactions determined by the Morse potential. 

Considering the specific parameters for  C$-$C bond one can estimate the minimum crossover energy to the chaotic state as $E_{c,q} \approx 0.7 E_{d}$ and transition to the quantum regime  is expected at $N>N_{c} \approx 20$ atoms. 

It is interesting to find out how long the chain of carbons should be  to attain the chaotic state at room temperature. Since in the Morse potential model room temperature is much smaller than the characteristic quantization energy $\hbar\sqrt{k/M} \sim 1300$cm$^{-1}$ we should use the quantum expression for the energy $E \approx N\pi^2 (k_{B}T)^2/(\hbar\sqrt{k/M})$. Setting $E \sim E_{c,q}$ and $k_{B}T \sim 4\cdot 10^{-21}$J at room temperature we get the estimate 
\begin{eqnarray}
N_{C}=\frac{4.2E_{d}\hbar\sqrt{k/M}}{\pi^2(k_{B}T)^2}=425. 
\label{eq:NminCrt}
\end{eqnarray}
This number is very large. 

The energy of molecules studied in the experiments \cite{Stewart83} is of order of $3000$cm$^{-1}$, which is much lower than the minimum energy $E_{c,q}$ needed to reach the chaotic state; yet some of them show a fast internal relaxation.  Therefore, a Morse potential based model of the FPU atomic chain seems to be not quite relevant there. Perhaps this is because real molecules (e. g.  alkane chains) are not perfectly linear but have a zig-zag shape making them much softer. Also transverse and optical modes have been ignored, while their effect can be significant \cite{LoganWolynes90,BigwoodLeitner98}. Accurate studies of molecules thus require more accurate definitions of their parameters.

\section{Materials and Methods}
\label{sec:Methods}

Analytical estimates use the analysis of resonant interactions. These methods can be qualitatively justified by the similarity of the problem to the exactly solvable localization problem on the Bethe lattice \cite{LoganWolynes90,ab16preprintSG,abGorniyMirlinDot}. We ignore logarithmic factors appearing in these considerations being concentrated on the power law dependencies. 

The numerical study exploits the exact diagonalization of Hamiltonian matrices using the standard MatLab software facilities \cite{MATLAB:2017b}. 

\section{Conclusion}
\label{sec:concl}

Here we briefly summarize the results of the present work. The semi-quantitative theory is developed to determine  the critical energy separating localized (integrable) and chaotic behaviors in the quantum FPU chain of atoms with different boundary conditions. The criterion of delocalization has been suggested considering resonant interactions for combined $\alpha+\beta$ FPU problem. It is predicted that the critical energy decreases with the number of atoms inversely proportionally to this number until the effective thermal energy exceeds the normal mode quantization energy in agreement with previous analysis of the classical $\beta$ FPU problem. At larger number of atoms the critical energy does not depend on this number.

The attempt of numerical verification of the results has been made in the oversimplified model with conserving number of phonons. This model shows that the chaos emerges at smaller energies for free and fixed ends boundary conditions compared to the system with periodic boundary conditions because of the smaller phase space in the latter case. The behaviors obtained  are consistent with theory predictions but the more realistic models need to be studied for the accurate theory verification. 

The application of the theory to atomic chains of carbon atoms described by the Morse potential predicts the occurrence of chaotic behavior for very long chains and high system energy that does not agree with experimental observations. Most probably this is because the model describes perfectly linear chains, while realistic (e. g. alkane) chains have more complicated structure and should be modeled with modified parameters.




\funding{This research was funded by NSF (CHE-1462075) and the Tulane  Bridge Fund.}

\acknowledgments{Authors acknowledges Sergei Flach, Ivan Khaymovich and Igor Rubtsov for stimulating discussion.}

\conflictsofinterest{The authors declare no conflict of interest. The funders had no role in the design of the study; in the collection, analyses, or interpretation of data; in the writing of the manuscript, or in the decision to publish the results.} 

\abbreviations{The following abbreviations are used in this manuscript:\\

\noindent 
\begin{tabular}{@{}ll}
FPU & Fermi-Pasta-Ulam
\end{tabular}}


\externalbibliography{yes}
\bibliography{Vibr}



\end{document}